\documentclass[a4paper,fleqn]{cas-dc}
\usepackage{graphicx, float}
\usepackage{amsmath}
\usepackage{mathrsfs}
\usepackage[utf8]{inputenc}
\usepackage{caption}
\usepackage[position=top]{subcaption}
\usepackage{comment}
\usepackage{bbding}
\usepackage[export]{adjustbox}
\usepackage{standalone}
\usepackage{rotating}
\usepackage[T1]{fontenc} 
\usepackage{utfsym}

\usepackage[per-mode=symbol]{siunitx}
\usepackage[normalem]{ulem}
\usepackage{cancel}
\usepackage[dvipsnames]{xcolor}
\usepackage[nameinlink,noabbrev]{cleveref}
\crefname{equation}{equation}{equations}

\usepackage{bm}
\usepackage{tikz}
\usetikzlibrary{decorations.pathreplacing}
\usepackage{breakcites}
\usetikzlibrary{shapes}
\usepackage{mathtools}
\definecolor{red2}{RGB}{192, 16, 16}
\definecolor{green2}{RGB}{48, 128, 48}
\definecolor{green3}{RGB}{0, 200, 0}
\definecolor{blue2}{RGB}{16, 80, 160}
\definecolor{blue3}{RGB}{0, 0, 255}
\definecolor{orange2}{RGB}{192, 128, 64}
\definecolor{yellow2}{RGB}{224, 192, 32}
\definecolor{purple2}{RGB}{96, 64, 160}
\definecolor{teal2}{RGB}{32, 112, 144}
\usepackage{xcolor}
\definecolor{drkg}{RGB}{105,105,105}
\definecolor{midg}{RGB}{169,169,169}
\definecolor{lgtg}{RGB}{220,220,220}
\definecolor{drkb}{RGB}{0,25,51}
\definecolor{lgtb}{RGB}{0,102,204}
\definecolor{lightg}{RGB}{160,160,160}
\definecolor{lightgr}{RGB}{30,120,30}
\usepackage{color}

\newcommand{\YY}{Y_n^m}

\usepackage[numbers]{natbib}
\usepackage{comment}

\def\tsc#1{\csdef{#1}{\textsc{\lowercase{#1}}\xspace}}
\tsc{WGM}
\tsc{QE}
\tsc{EP}
\tsc{PMS}
\tsc{BEC}
\tsc{DE}
\begin{document}
\let\WriteBookmarks\relax
\def\floatpagepagefraction{1}
\def\textpagefraction{.001}
\shorttitle{Microbubble surface instabilities in a strain stiffening viscoelastic material}
\shortauthors{S. Remillard et~al.}

\title [mode = title]{Microbubble surface instabilities in a strain stiffening viscoelastic material}                      
% \tnotemark[1,2]

%\tnotetext[1]{This document is the results of the research
 %  project funded by the National Science Foundation.}

%\tnotetext[2]{The second title footnote which is a longer text matter
 %  to fill through the whole text width and overflow into
 %  another line in the footnotes area of the first page.}

\author[1]{Sawyer Remillard}[orcid=0000-0002-5469-6168]
\cormark[1]
\credit{Conceptualization, Data Curation, Formal Analysis, Investigation, Methodology, Software, Validation, Writing - Original draft preparation, Writing - review \& editing}
\author[2]{Bachir A. Abeid}[orcid=0009-0009-0859-8434]
\credit{Data Curation, Investigation, Methodology,  Writing - review \& editing}
\author[3]{Timothy L. Hall}[orcid=0000-0002-4867-2856]
\credit{Methodology, Resources, Writing - review \& editing}
\author[3]{Jonathan R. Sukovich}[orcid=0000-0002-5650-991X]
\credit{Conceptualization, Methodology, Resources, Writing - review \& editing}
\author[4]{Jacob Baker}[orcid=0000-0003-3361-1513]
\credit{Writing - Original draft preparation, Writing - review \& editing}
\author[5,6]{Jin Yang}[orcid=0000-0002-5967-980X]
\credit{Conceptualization, Data curation, Funding Acquisition, Writing - review \& editing}
\author[2]{Jonathan B. Estrada}[orcid=0000-0003-1083-4597]
\credit{Conceptualization, Funding Acquisition, Investigation, Methodology, Resources, Supervision, Writing - review \& editing}
\author[1]{Mauro Rodriguez Jr.}[orcid=0000-0003-0545-0265]
\cormark[2]
\credit{Conceptualization, Funding Acquisition, Investigation, Methodology, Project Administration, Resources, Supervision, Writing - original draft, Writing - review \& editing}

\affiliation[1]{organization={School of Engineering, Brown University}, 
                city={Providence},
                postcode={02912}, 
                state={RI},
                country={USA}}

\affiliation[2]{organization={Department of Mechanical Engineering, University of Michigan},
                city={Ann Arbor},
                postcode={48109}, 
                state={MI},
                country={USA}}

\affiliation[3]{organization={Department of Biomedical Engineering, University of Michigan},
                city={Ann Arbor},
                postcode={48109}, 
                state={MI},
                country={USA}}

\affiliation[4]{organization={Department of Biomedical Engineering, The University of Texas at Austin},
                city={Austin},
                postcode={78712}, 
                state={TX},
                country={USA}}
\affiliation[5]{organization={Department of Aerospace Engineering and Engineering Mechanics, The University of Texas at Austin},
                city={Austin},
                postcode={78712}, 
                state={TX},
                country={USA}}
\affiliation[6]{organization={Texas Materials Institute, The University of Texas at Austin},
                city={Austin},
                postcode={78712}, 
                state={TX},
                country={USA}}

\cortext[cor1]{Corresponding author}
\cortext[cor2]{Principal corresponding author}
%\fntext[fn1]{This is the first author footnote, but is common to third
%  author as well.}
%\fntext[fn2]{Another author footnote, this is a very long footnote and
%  it should be a really long footnote. But this footnote is not yet
  %sufficiently long enough to make two lines of footnote text.}

%\nonumnote{This note has no numbers. In this work we demonstrate $a_b$
%  the formation Y\_1 of a new type of polariton on the interface
  %between a cuprous oxide slab and a polystyrene micro-sphere placed
  %on the slab.
  %}

\begin{abstract}
Understanding the dynamics of instabilities along fluid-solid interfaces is critical for the efficacy of focused ultrasound therapy tools (e.g., histotripsy) and microcavitation rheometry techniques. 
Non-uniform pressure fields generated by either ultrasound or a focused laser can cause non-spherical microcavitation bubbles. 
Previous perturbation amplitude evolution models in viscoelastic materials either assume pure radial deformation or have inconsistent kinematic fields between the fluid and solid contributions. 
We derive a kinematically-consistent theoretical model for the evolution of surface perturbations. 
The model captures the non-linear kinematics of a strain-stiffening viscoelastic material surrounding a non-spherical bubble. 
The model is validated for (i) small, approximately linear radial oscillations and (ii) large inertial oscillations using laser-induced microcavitation experiments in a soft hydrogel. 
For the former, the bubble is allowed to reach mechanical equilibrium, and then surface perturbations are excited using ultrasound forcing. 
For the latter, the microbubble forms small bubble surface perturbations at its maximum radius that grow during collapse. 
The model's dominant surface perturbation mode scales linearly with equilibrium radius and matches experiments.
Similarly, the model's perturbation amplitude evolution sufficiently constrains the rheometry problem and is experimentally validated.
\end{abstract}

% \begin{graphicalabstract}
% \includegraphics{cas-grabs.pdf}
% \end{graphicalabstract}

\begin{highlights}
\item Kinematically-consistent bubble surface perturbation evolution model for strain stiffening viscoelastic materials
\item Theoretical model validation for the evolution of bubble surface perturbations in viscoelastic materials
\item Tunable strain-rate material characterization from a single experimental setup
\end{highlights}


\maketitle

\section{Introduction}
Experiments and simulations show that cavitation bubbles rarely stay perfectly spherical during their oscillations, especially in soft viscoelastic materials \citep{JOHNSEN_COLONIUS_2009, Brujan_vogel_2006, Hamaguchi2015, Lauterborn_Bolle_1975, Plesset1977, Murakami2020, Murakami2021, Yang2021, yang2021dynamic-ruga, yang2022mechanical}. 
Asphericities form during applications of cavitation in soft materials such as histotripsy \citep{Vlaisavljevich2015a, Mancia2017a, MAXWELL20091982, Xu2007, Bailey2003} and Inertial Microcavitation Rheometry (IMR) \citep{Prosperetti1977, Estrada2018, YANG2020} which can decrease their efficacy.
In contrast, the non-spherical oscillations increase the efficacy of targeted drug release \citep{Lajoinie2018, Tinkov2010,Kheirolomoom2007}.
However, modeling this phenomenon is limited to two models across three studies \citep{Yang2021, Gaudron2020, Murakami2020} where each model uses different assumptions, leading to disagreement.

Although the study of non-spherical cavity oscillations in viscoelastic solid materials is quite recent, it has been a topic studied extensively in fluids, dating back to \citet{Plesset1954}.
\citet{Prosperetti1977} included viscosity and vorticity effects.
The latter led to a computationally expensive model, which must be solved directly rather than analytically linearized for a stability analysis.
%\mrdz{I am not sure what you are trying to say in this previous sentence}
To simplify Prosperetti's model, \citet{Hilgenfeldt1996} restricted the vorticity modeling to a viscous boundary layer.
This model agreed with experimental data by \citet{Versluis2010}.
Later models include higher-order terms in the perturbation series expansion to compute energy transfer between the volumetric and the shape mode oscillations \citep{Harkin2013,Shaw2006,Shaw2025JMF,Shaw2025Pof}.
Variants of the model from \citet{Shaw2006} agreed with the following two experimental studies.
\citet{Claude2017} showed the coupling between the perturbation modes and radial dynamics, where ultrasound forcing was used over prolonged periods of time to excite these non-spherical modes.
In the work by \citet{Philippe2018}, nonspherical oscillations were first induced through bubble coalescence, then forced via ultrasound transducers.
However, these models are only valid for Newtonian fluids, a special case of a material model whose viscous Cauchy stress is proportional to the applied strain rate.

Non-spherical cavity modeling has been extended to elastic solid materials. 
\citet{Gaudron2020} developed a model of non-spherical bubble oscillations and considered both neo-Hookean hyperelastic and linear elastic material models. 
The model was extended by \citet{Murakami2020} to include vorticity and viscosity and was shown to agree with part of the experimental data from \citet{Hamaguchi2015} as well as \citet{Garbin2020}.
However, the models of \citet{Murakami2020} and \citet{Gaudron2020} utilized inconsistent kinematic fields for acceleration and material deformation.
The velocity potential approach from \citet{Plesset1954} was used for the former, while for the latter, when computing the elastic contribution to the divergence of the Cauchy Stress, pure radial deformation was assumed.
Using two different kinematic fields on the same material is physically unrealistic.
The approach leads to incorrect predictions when the shear modulus becomes large or if the material experiences large stretch ratios.
Separately, \citet{Yang2021} developed a nonspherical model for visco-hyperelastic materials which show strain stiffening.
They show agreement with experiments for the model's prediction of the instability growth and decay through an eigenvalue analysis of the model governing equations. 
The model uses consistent kinematic fields throughout the derivation, however, only considers radial motion and neglects angular deformations. 
To date, the non-spherical models of \citep{Yang2021, Gaudron2020, Murakami2020} which account for viscoelasticity have not been validated with perturbation evolution data from experiments.

We hypothesize that angular deformation is necessary for comparing direct observations of the perturbation mode and amplitude evolution to the theory as it was considered for fluid validated models.
Moreover, non-spherical oscillations can be used to characterize viscoelastic materials across a range of strain rates.
For an ultrasound forced system, the strain rate experienced by the material is tunable by the forcing frequency and bubble size.
To that end, the objective of this work is to: (i) develop and validate a kinematically consistent, spherical surface perturbation model that includes angular deformations and (ii) use perturbation dynamics to constrain material characterization techniques.
The proposed model is presented in \cref{sec:theoretical_model}.
Small radial oscillations model simplifications are presented in \cref{sec:stability}.
The model is compared with existing models in \cref{sec:model_summary_comparison}.
The numerical and experimental methods are described in \cref{sec:methods}. 
The results and discussion comparing theory and experiments are presented in \cref{sec:results}.
We conclude with our contributions in \cref{sec:conclusions}.

\section{Kinematically consistent analytical model}
\label{sec:theoretical_model}
\subsection{Geometry}
\begin{figure*}
    \centering
    \includegraphics[width=\linewidth]{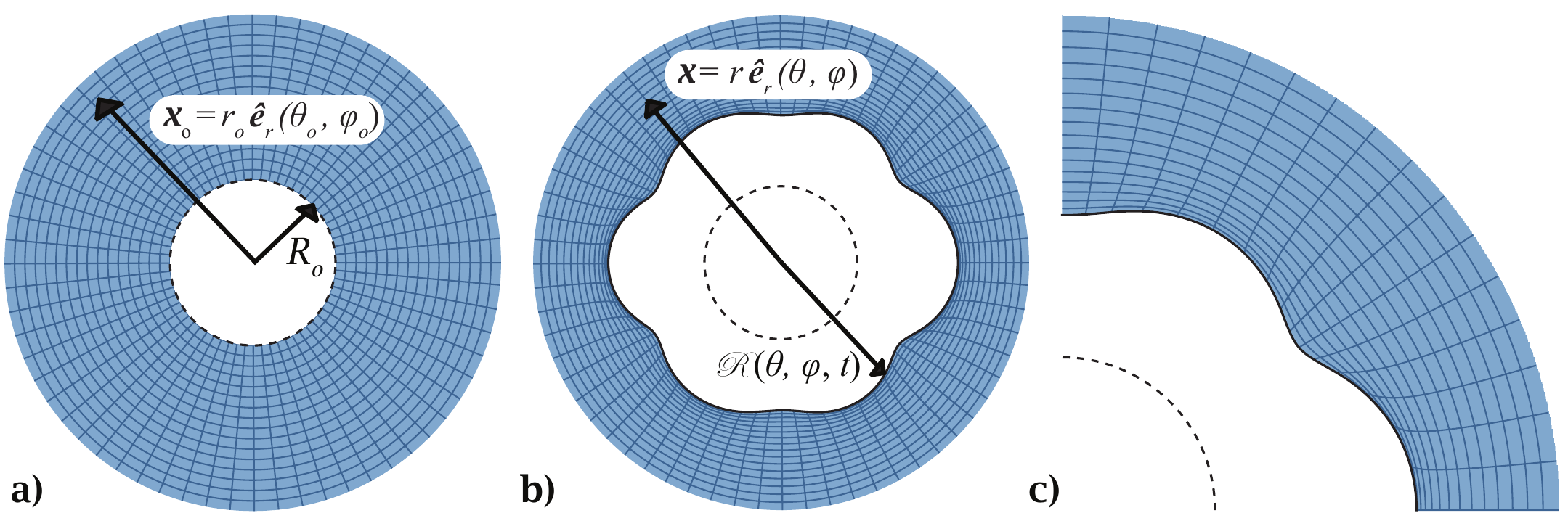}
    \caption{Problem set-up shown as a cross-section for constant $\phi$. 
    \textbf{a)} Reference configuration with the bubble at the mechanical equilibrium radius shown in dashed lines. 
    \textbf{b)} An example current configuration with base radial growth and formation of the spherical harmonic instabilities. 
    \textbf{c)} Zoom in of material grid and bubble distortion of the top right quadrant of b).}
    \label{fig:geometric_setup}
\end{figure*}

\Cref{fig:geometric_setup} shows the problem set-up and an example of a uniform grid transformation from the reference configuration from axisymmetric spherical harmonic modes.
A spherical bubble is in stress-free mechanical equilibrium with the surrounding material and reference radius $R_o$.
The bubble surface is
\begin{equation*}
    \mathscr{R}(t, \theta, \phi)=R(t)\left(1+\sum_{n,m}\epsilon_n(t) \YY\right),
\end{equation*} 
where $R$ is the mean bubble radius, $\epsilon_n$ the non-dimensional amplitude of the surface perturbation, $t$ time, and $Y_n^m = \YY(\theta,\phi)$ the real spherical harmonics with mode number $n$ and azimuthal order $m$ in the current coordinates.
Perturbation amplitude and bubble radius time dependency and summation of the modes are dropped for brevity.

The reference spherical coordinate system is $\psi_o = \left(r_o, \theta_o, \phi_o\right)$ and the current $\psi = \left(r, \theta, \phi \right)$.
The position vectors in the reference and current coordinates are, 
\begin{equation}
    \mathbf{x}_o = r_o \, \hat{\mathbf{e}}_r(\theta_o, \phi_o), \quad \mathbf{x} = r \, \hat{\mathbf{e}}_r(\theta, \phi),
    \label{eq:position}
\end{equation}
where $\mathbf{\hat{e}}_r$ is the radial direction unit vector.
The current position vector is decomposed into the base radial deformation with superposed first order small non-spherical deformation, 
\begin{equation}
    \mathbf{x} = r_s\hat{\mathbf{e}}_r(\theta_o,\phi_o)+\epsilon_n\mathbf{u}_1,
    \label{eq:small_on_large}
\end{equation}
where $r_s$ is the base spherical deformation, and $\mathbf{u}_1$ the nonspherical surface perturbation deformation vector.
Since the non-spherical motion is assumed to be small relative to the spherical motion, i.e., $\epsilon_n(t) \ll 1$, then terms $\mathscr{O}(\epsilon_n^2)$ and higher are truncated.
Then, the motion map from the reference to current coordinates is denoted as $\boldsymbol{\chi}$:
\begin{equation*}
    \psi = \boldsymbol{\chi}\left(\psi_o \right) 
    = \boldsymbol{\chi}_s\left( \psi_o\right) + \epsilon_n\boldsymbol{\chi}_1\left(\psi_o \right),
\end{equation*}    
where,
\begin{equation*}
    \begin{aligned}
        \boldsymbol{\chi}_s\left( \psi_o\right)& =  \left(r_s(r_o, t), \, \theta_o, \, \phi_o\right),\\
        \boldsymbol{\chi}_1\left( \psi_o\right) &= \left(r_1(\psi_o, t), \,  \theta_1(\psi_o, t),\,  \phi_1(\psi_o, t)\right).
    \end{aligned}
\end{equation*}
Here, $\chi_1$, $r_1$, $\theta_1$, and $\phi_1$ are the respective first-order terms to the total coordinate, radial, $\theta$, and $\phi$ maps.
Plugging the coordinate deformation map into \cref{eq:position,eq:small_on_large} and linearizing we obtain 
\begin{equation}
    \mathbf{u}_1 = r_1\hat{\mathbf{e}}_r + r_s\theta_1\hat{\mathbf{e}}_\theta + r_s\sin\theta_o \phi_1 \hat{\mathbf{e}}_\phi,
    \label{eq:pert_disp}
\end{equation}
where $\hat{\mathbf{e}}_\theta$ and $\hat{\mathbf{e}}_\phi$ are unit vectors in the polar and azimuthal direction, respectively.
The unit vectors in \cref{eq:pert_disp} are functions of $\theta_o$ and $\phi_o$ only.
Since $(\theta,\phi) = (\theta_o, \phi_o)+\mathscr{O}(\epsilon_n)$, then when computing the position vector, the components of $\mathbf{u}_1$ can be written in either coordinate basis to first order accuracy.

%a one-to-one mapping between the reference and current configurations.

\subsection{Kinematics}
\label{sec:kinematics}

%The push-forward representation of the displacement vector is 
%\begin{equation*}
%    \mathbf{u} = \mathbf{H} (\mathbf{x} - \mathbf{x}_o)= [r-r_o, \, r(\theta-\theta_o), \, r\sin\theta(\phi-\phi_o)]^{\top} = \mathbf{u}_s+ \epsilon_n\mathbf{u}_1,
%\end{equation*} 
%where $\mathbf{H}$ is the diagonal matrix of the lame scale factors, $(1,r,r\sin\theta)$ and 0 on the off diagonal.
%The respective base and perturbed state displacement fields are 
%\begin{equation*}
%    \mathbf{u}_s = [r_s-r_o, \, 0, \, 0]^{\top}, \quad \mathbf{u}_1 = [r_1, \, r_s\theta_1, \, r_s \phi_1 \sin(\theta_o) ]^{\top}.
%    \label{eq:pert_disp_vec}
%\end{equation*}
The deformation gradient tensor is
\begin{equation}
\begin{aligned}
& \mathbf{F}=\frac{\partial \mathbf{x}}{\partial \mathbf{x}_o} =
 \left[\begin{array}{ccc}
 \frac{\partial r}{\partial r_o} & \frac{1}{r_o}\frac{\partial r}{\partial \theta_o} & \frac{1}{r_o \sin\theta_o}\frac{\partial r}{\partial \phi_o} \\[4pt]
 r\frac{\partial \theta}{\partial r_o} & \frac{r}{r_o}\frac{\partial \theta}{\partial \theta_o} & \frac{r}{r_o \sin\theta_o}\frac{\partial \theta}{\partial \phi_o} \\[4pt]
 r \sin\theta\frac{\partial \phi}{\partial r_o} & \frac{r \sin\theta}{r_o}\frac{\partial \phi}{\partial \theta_o} & \frac{r \sin\theta}{r_o\sin\theta_o} \frac{\partial \phi}{\partial \phi_o}
\end{array}\right],
\label{eq:full_def_grad}
\end{aligned}
\end{equation}
and written in the expanded coordinate maps in \cite{supplementary_information}.
The deformation gradient can be decomposed as a sum of powers of $\epsilon_n$, $\mathbf{F} = \mathbf{F}_s+\epsilon_n \mathbf{F}_1$.
Material incompressibility is satisfied with $\det{(\mathbf{F})} = 1$.
After truncating terms of $\mathscr{O}(\epsilon_n^2)$ and higher, the incompressibility condition becomes
\begin{equation*}
    \text{det}(\mathbf{F}) \approx \text{det}(\mathbf{F}_s)+\epsilon_n \left[\frac{\partial}{\partial \epsilon_n} \text{det}(\mathbf{F})\right]_{\epsilon_n = 0} = 1.  
\end{equation*}
By the method of perturbation matching and consistent with \citet{Ogden1984}, $\det{(\mathbf{F}_s)} = 1$ and the incremental deformation incompressibility condition is
\begin{equation}
    \left[\frac{\partial}{\partial \epsilon_n} \text{det}(\mathbf{F})\right]_{\epsilon_n = 0}  = \text{tr}\left(\mathbf{F}_s^{-1}\mathbf{F}_1 \right)=\nabla_{\mathbf{x}}\cdot \mathbf{u}_1=0.
    \label{eq:lin_incomp}
\end{equation}
The base state deformation for the purely spherical case is well known as \citep{Estrada2018, Gaudron2015,Yang2021}.
The incompressibility relationship for the base deformation is
\begin{equation}
        r_s  = \left(r_{o}^3+ R^3 -R_o^3\right)^{1/3}.
\end{equation}
\Cref{eq:lin_incomp} is solved for the mappings proportional to $\epsilon_n$. 
We consider the following ansatz for the displacement,
\begin{equation}
    \mathbf{u}_1 = \nabla_{\mathbf{x}}\left[f(r) \YY\right],
    \label{eq:ansatz}
\end{equation}
where $f(r)$ is a function of the spatial radial coordinate.
The rotational part of the perturbed displacement field which would form due to the presence of interfacial perturbations, viscosity, and elasticity is not considered for simplicity.

Substituting \cref{eq:ansatz} into the spatial divergence of the perturbed displacement field, linearizing, and using the characteristic equation of the spherical harmonics, the incompressibility condition simplifies to an ordinary differential equation
\begin{equation}
    r^2f''+2rf'-n(n+1)f = 0.
\end{equation} 
The corresponding decaying solution of the above equation is $f = \mathcal{C}_1 r^{-n-1}$, where $\mathcal{C}_1$ is solved for via the radial displacement boundary condition on the bubble surface:
\begin{equation*}
    \begin{aligned}
    r(r_o= R_o, \theta_o,\phi_o,t) = R(1+\epsilon_n \YY ) \to
    f'|_{r = R}\YY = R \YY,
    \end{aligned}
\end{equation*}
with $\mathcal{C}_1 = -R^{n+3}/(n+1)$.
Then, the respective velocity and acceleration are calculated via,
\begin{subequations}
    \begin{equation}
        \mathbf{v} = \dot{r} \, \hat{\mathbf{e}}_r + r\dot{\theta} \, \hat{\mathbf{e}}_{\theta}+ r\sin\theta \dot{\phi} \, \hat{\mathbf{e}}_{\phi},
        \label{eq:velocity}
    \end{equation}
    \begin{equation}
        \begin{aligned}
            \boldsymbol{a} &= \left(\ddot{r}-r\dot{\theta}^2-r\dot{\phi}^2  \sin^2\theta \, \right)\hat{\mathbf{e}}_r \\
            &+ \left(r\ddot{\theta}+ 2\dot{r}\dot{\theta}-r\dot{\phi}^2\sin\theta \cos\theta \right) \hat{\mathbf{e}}_{\theta} \\
            &+ \left(r\ddot{\phi}  \sin \theta +2\dot{r}\dot{\phi}\sin\theta+ 2 r \dot{\theta}\dot{\phi}\cos \theta\right) \hat{\mathbf{e}}_{\phi}, 
        \end{aligned}
    \end{equation}
\end{subequations}
where dot operators are time derivatives at constant reference coordinates and the unit vectors are with respect to the current coordinate basis.
The coordinate deformation map, linearized inverse coordinate deformation map, and Eulerian velocities and accelerations are listed in \cite{supplementary_information}.

\subsection{Momentum balance}
Momentum balance equation is
\begin{equation}
    \rho \boldsymbol{a} = \nabla_{\mathbf{x}} \cdot\boldsymbol{\sigma},
    \label{eq:mom_balance}
\end{equation}
where $\rho$ is the material density and $\boldsymbol{\sigma}$ the Cauchy stress tensor. 
Considering the quadratic Kelvin-Voigt (qKV) constitutive model \citep{Knowles1977, YANG2020},
\begin{equation}
    \boldsymbol{\sigma} = G\left(1+\alpha(I_B-3) \right)\mathbf{B}+2\mu \mathbf{D}-\mathcal{P}\mathbf{I},
    \label{eq:const_model}
\end{equation}
where $G$ is the shear modulus, $\alpha$ the strain stiffening parameter,  $\mu$ the viscosity, and $\mathbf{I}$ the identity tensor.
$I_B$ is the first invariant of $\mathbf{B}$, linearized with respect to the perturbation amplitude, i.e.,
\begin{equation}
\begin{aligned}
    I_B &=   \frac{2 r^6+\varrho^2}{r^4\varrho^{2/3}}+  \frac{ 2\epsilon_n \, R^{n+3} }{r^{n+7}\varrho^{5/3}}\left(R^3-R_o^3\right) \left(2 r^3-R^3+R_o^3\right) \\
    & \times\left((n+2) r^3+n(R_o^3-R^3)\right)\YY,
\end{aligned}
\end{equation}
where $\varrho = \left(r^3-R^3+R_o^3 \right)^{1/3}$.
$\mathbf{B} = \mathbf{F} \mathbf{F}^{\top}$ is the left Cauchy-Green deformation tensor and not deviatoric, thus $\mathcal{P}$ is a pressure-like quantity and related to the hydrostatic pressure.
$\mathbf{D}$ is the rate of strain tensor, i.e.,
\begin{equation*}
    \mathbf{D} = \frac{1}{2}\left(\nabla_{\mathbf{x}}\mathbf{v}+(\nabla_{\mathbf{x}}\mathbf{v})^{\top} \right).
\end{equation*}
The components of the $\mathbf{B}$ and $\mathbf{D}$ tensors used in \cref{eq:const_model} are listed in \cite{supplementary_information}.
Additionally, 
\begin{equation*}
    \mathcal{P} = \mathcal{P}_s+ \epsilon_n \mathcal{P}_1\YY,
\end{equation*}
where $\mathcal{P}_s$ and $\mathcal{P}_1$ are the respective base state pressure and  surface perturbations contributions.

\subsection{Boundary conditions}
The two boundary conditions of the system are (i) traction at the bubble wall and (ii) stress-free in the far field.
The inertia and viscosity of the gas or vapor is negligible compared to that of the surrounding fluid \citep{Brennen1995,Murakami2020}.
After linearization, the radial-normal component of the traction and far field boundary condition are
\begin{subequations}
    \begin{equation}
        \sigma_{rr} |_{r = R(1+\epsilon_n \YY)}= -p_{\text{b}} + \frac{2\gamma}{R} + \frac{(n+2)(n-1)\gamma}{R}\epsilon_n \YY,
    \end{equation}
    \begin{equation}
        \left . \sigma_{rr}\right|_{r\rightarrow\infty} = -p_{\infty},
    \end{equation}
\end{subequations}
respectively, where $\gamma$ is the surface tension between the two phases, $p_{\text{b}}$ the bubble internal pressure, $p_{\infty}$ the far-field pressure.
Similar to \cite{Gaudron2015, Murakami2020, Yang2021}, we lack the degrees of freedom in our ansatz to enforce tangential stress continuity. 

\subsection{Equations of motion}
Integrating the radial component of \cref{eq:mom_balance}, applying spherical orthogonality (multiplying by $\sin (\theta)Y_0^0$ and integrating in $\theta$ and $\phi$), and applying the two stress boundary conditions yields the Rayleigh-Plesset equation,
\begin{equation}
    R\ddot{R} + \frac{3}{2}\dot{R}^2 = \frac{1}{\rho_{\text{m}}}\left(p_{\text{b}}-p_{\infty}-\frac{2\gamma}{R}+S \right).
    \label{eq:RP}
\end{equation}
The stress integral for a qKV model is
\begin{equation}
\begin{aligned}
    S&=  \frac{(3 \alpha-1) G}{2}\left[5-\lambda^{-4}-4\lambda^{-1}\right] \\
    &  +2 \alpha G\left[\frac{27}{40}+\frac{1}{8}\lambda^{-8}+\frac{1}{5}\lambda^{-5}+\lambda^{-2}-2\lambda\right] -\frac{4 \mu \dot{R}}{R}, 
\end{aligned}
\end{equation}
where $\lambda$ is the spherical stretch ratio of the material at the bubble wall $\lambda = R/R_o$.
Multiplying \cref{eq:mom_balance} by $\sin (\theta)\YY$ and applying orthogonality yields:
\begin{equation}
\ddot{\epsilon}_n + \eta \dot{\epsilon}_n+ \xi\epsilon_n = 0,
\label{eq:pert_evo}
\end{equation}
where coefficients $\eta$ and $\xi$ for the current model and models of \citet{Yang2021} and \citet{Murakami2020} without the rotational corrections are in~\cite{supplementary_information}.
Since $\mathscr{O}(\epsilon_n^2)$ and higher terms are neglected, modes are decoupled and depend only on radial motion.
Here, we only consider the irrotational models of \citet{Murakami2020} and \citet{Prosperetti1977}.

\section{Linearized stability analysis}
\label{sec:stability}
Predicting the initial condition to the growth of the surface perturbations under ultrasonic forcing is challenging in experiments because it often begins as a seed of the order of $\mathscr{O}(\SI{E-3}{})$~\citep{Murakami2020}.
Upon termination of the ultrasonic forcing, the perturbation amplitudes decay to equilibrium. 
The initial condition for the decay process is simpler since the perturbation amplitudes will be of finite amplitude.
Considering small radial oscillations due to ultrasonic forcing, the bubble radius evolution can be approximated as $R(t) = R_o(1+\Delta \lambda(t))$ where $\Delta \lambda(t) = \lambda(t)-1  = \mathscr{O}(\epsilon_n)$.
The pressure surrounding the bubble is $p_\infty = p_{\rm atm} + \mathscr{A}\cos(2 \pi f t)$, where $\mathscr{A}$ is the amplitude of the forcing and $f$ the frequency.
Linearizing \cref{eq:pert_evo} about small radial oscillations, the  coefficients are
\begin{subequations}
\begin{equation}
    \eta_{\rm l} = \frac{2 (n+2)(n+1) \mu }{ \rho R_o^2},
\end{equation}
\label{eq:lin_coeff_eta}
\begin{equation}
    \xi_{\rm l} =\omega_n^2 = \frac{2(n+2) (n+1)}{\rho  R_o^2}\left(G + \frac{(n-1) \gamma}{2 R_o}\right), 
\end{equation}%
\label{eq:lin_coeff_xi}%
\end{subequations}%
where the subscript $\rm l$ denotes linearized quantities and $\omega_n$ is the surface perturbation mode natural frequency.

The prediction of the most unstable (dominant) mode for given material properties, ultrasound forcing, and equilibrium bubble radius satisfy the parametric resonance condition, $2\omega_n/\omega_0=1$,
where $\omega_0 = 2\pi f$ is the angular forcing frequency.
This 1:2 resonance condition is from Mathieu's equation theory \cite{Matsumoto_2012} and was verified experimentally for non-spherical bubbles \cite{Versluis2010}.

Prescribing an initial finite amplitude, $\epsilon_{n,0} = \epsilon_n(t=0)$, and $\dot{\epsilon}_n(t=0) = 0$, there is an analytical solution for the perturbation equation for linearized radial motion:
\begin{equation}
    \epsilon_n(t) = \epsilon_{n,0} \text{exp}\left(-\frac{t \eta_{\rm l}}{2}\right) \Bigg[\cosh\left(\frac{t \delta_{\text{l}}}{2} \right) 
    + \frac{\eta_{\rm l}}{\delta_{\text{l}}} \sinh \left(\frac{t \delta_{\text{l}}}{2} \right) \Bigg], 
    \label{eq:pert_evo_linear}
\end{equation}
where $\delta_{\text{l}} = \sqrt{\eta_{\rm l}^2-4\xi_{\rm l}}$.

\section{Comparison of analytical models}
\label{sec:model_summary_comparison}

\begin{figure*}
    \centering
    \includegraphics[width=\linewidth]{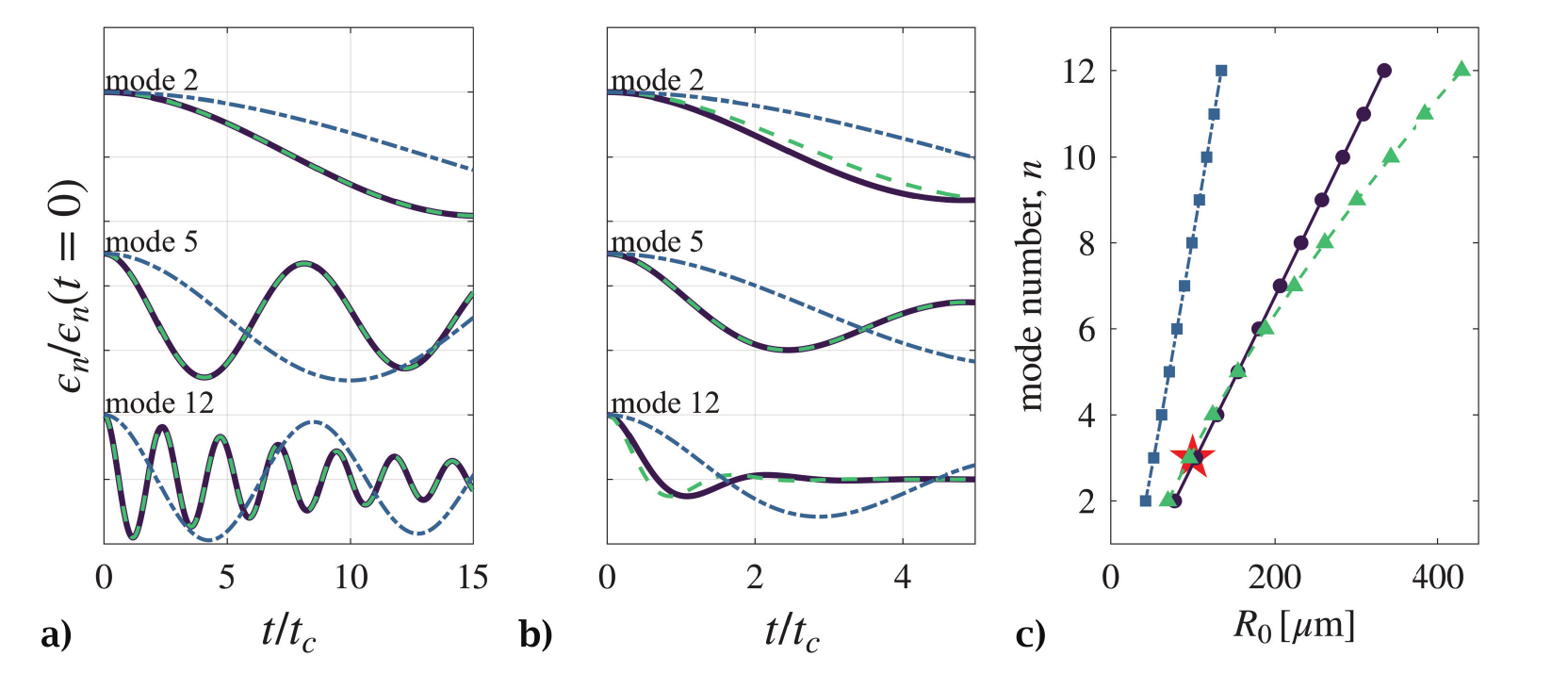}
    \caption{Perturbation evolution under linear radial motion for \textbf{a)} fluid \textbf{b)} a viscoelastic solid and \textbf{c)} most unstable mode prediction.
    Dashed green lines: \citet{Murakami2020}; dashed-dotted light blue lines: \citet{Yang2021};  and solid dark blue lines: current theory.}
    \label{fig:model_comp}
\end{figure*}
\Cref{tab:comparison} tabulates the differences between the models of \cite{Prosperetti1977}, \cite{Murakami2020}, \cite{Yang2021}, and the present model.
For zero shear modulus, the models from \citet{Murakami2020} and \citet{Prosperetti1977} are equivalent.
However, the model from \citet{Yang2021} and the present model do not simplify in this way for zero shear modulus.
Only in the limit of linearized radial oscillations and zero shear modulus, the current model and the model from \citet{Prosperetti1977} are equivalent, this is explained through vorticity.
The present perturbed displacement field is curl-free but does not guarantee a curl-free velocity or acceleration field.
Taking the Eulerian curl of the velocity described in \cref{eq:velocity}, we obtain:
\begin{equation}
    \nabla_{\mathbf{x}}\times \mathbf{v}=  6\epsilon_n\frac{n+2}{n+1}\frac{R^{5+n}\dot{R}}{r^{6+n}}\left(\frac{1}{\sin\theta}\frac{\partial \YY}{\partial \phi}\, \hat{\mathbf{e}}_{\theta} - \frac{\partial \YY}{\partial \theta}\, \hat{\mathbf{e}}_{\phi}\right).
\end{equation}
If the radial motion is linear or static, there is no vorticity and the irrotational model assumptions are valid \cite{Prosperetti1977}.
However, as is seen in \cite{supplementary_information}, this vorticity residual leads to a vorticity equation that we cannot directly enforce.
This is a persistent issue across the models of \citet{Yang2021} and \citet{Murakami2020} contributing to causing violation of the $\theta$ and $\phi$ components of the momentum balance.
To satisfy the three momentum balance equations, more degrees of freedom are needed in the ansatz for the displacement field.
However, it is shown by \citet{Prosperetti1977} that the dominant effect to the perturbation amplitude evolution (for viscous fluids), is due to the contribution from the gradient based ansatz.

Given linear radial motion and a zero-shear modulus, the coefficients, $\xi_{\rm l}$ and $\eta_{\rm l}$ from \citet{Prosperetti1977}, \citet{Murakami2020}, and the present work agree (see~\cite{supplementary_information}).
Comparing $\xi_{\rm l}$ and $\eta_{\rm l}$ from ~\citet{Yang2021}, \citet{Murakami2021}, and \cref{eq:lin_coeff_xi} and \cref{eq:lin_coeff_eta}, the coefficients of $\mu$ and $G$ are equivalent when the kinematic fields are consistent.
When kinematic fields are consistent, the leading order power of the equilibrium radius, $R_o$ matches the inverse of the leading order power of the mode number $n$ in $\xi_{\text{l}}$ which leads to a linear scaling of the most unstable mode as a function of $R_o$.
To show the equivalency between models, we consider a $\SI{200}{\micro\meter}$ bubble in water ($\mu = \SI{1}{\milli\pascal\cdot\second}$ and $\gamma = \SI{72}{\milli\newton\per\meter}$) with static radial motion.
We initialize modes 2, 5, and 12 out of equilibrium at an initial amplitude of $0.1$.
\Cref{fig:model_comp} a) shows the current model exactly matches the model by \citet{Prosperetti1977}, here $t_c = R_{\textrm{max}}\sqrt{\rho/p_{\infty}}$. 
The disagreement between the \citet{Yang2021} and present model is due to the missing factor of $\sqrt{n+1}$ in the perturbation mode natural frequency due to angular motion.
Inclusion of angular motion changes the solution of the displacement field obtained from the incompressibility condition to localize more material strain to the bubble surface, explaining the increased frequency.

With the same initial conditions we simulate a soft material with material properties calibrated in \citet{Hamaguchi2015}, $G = 1.7$ \SI{}{\kilo\pascal}, $\mu = 14.1$ \SI{}{\milli\pascal\cdot\second}, and $\gamma = \SI{40}{\milli\newton\per\meter}$.
The results are shown in \cref{fig:model_comp} b).
The model from \cite{Murakami2020} under and overestimates the frequency of oscillation and for low and high mode numbers, respectively.
The frequency discrepancy between the present model and the model by \citet{Murakami2020} is explained by the scaling of $\xi_{\rm l}$ with respect to the equilibrium radius and the mode number.
\Cref{fig:model_comp} c) shows the non-linear scaling between mode number and equilibrium radius predicted by \citet{Murakami2020}.
For $f = \SI{28}{\kilo\hertz}$, and $R_o \approx \SI{100}{\micro\meter}$, three is the most unstable mode observed by \cite{Hamaguchi2015}. 
We observe that the \citet{Murakami2020} and the present models correctly predict the third mode as the most unstable.
As equilibrium radii and therefore mode number increases, the difference between $\xi_{\rm l}$ from \citet{Murakami2020} and the present model diverges.

\begin{table}
    \centering
    \caption{Comparison of linearized bubble surface perturbation models.} 
    \begin{tabular}{c | c | c | c }
        Analysis & Angular &  Finite& Kinematic    \\ 
         & Motion & elastic  & consistency \\
         \hline
        \protect Prosperetti \cite{Prosperetti1977} & \Checkmark & $	\usym{2718}$ & \Checkmark \\ 
        \protect Murakami et al.\cite{Murakami2020} & Hybrid & \Checkmark & $\usym{2718}$\\ 
        \protect Yang et al. \cite{Yang2021} & $\usym{2718}$ & \Checkmark  & \Checkmark  \\ 
        Present work & \Checkmark & \Checkmark & \Checkmark\\
    \end{tabular}
     \label{tab:comparison}
\end{table}

\section{Methods\label{sec:methods}}

\subsection{Experiments}
\label{subsec:experimental_methods}
\begin{figure*}
    \centering
    \includegraphics[width=\textwidth]{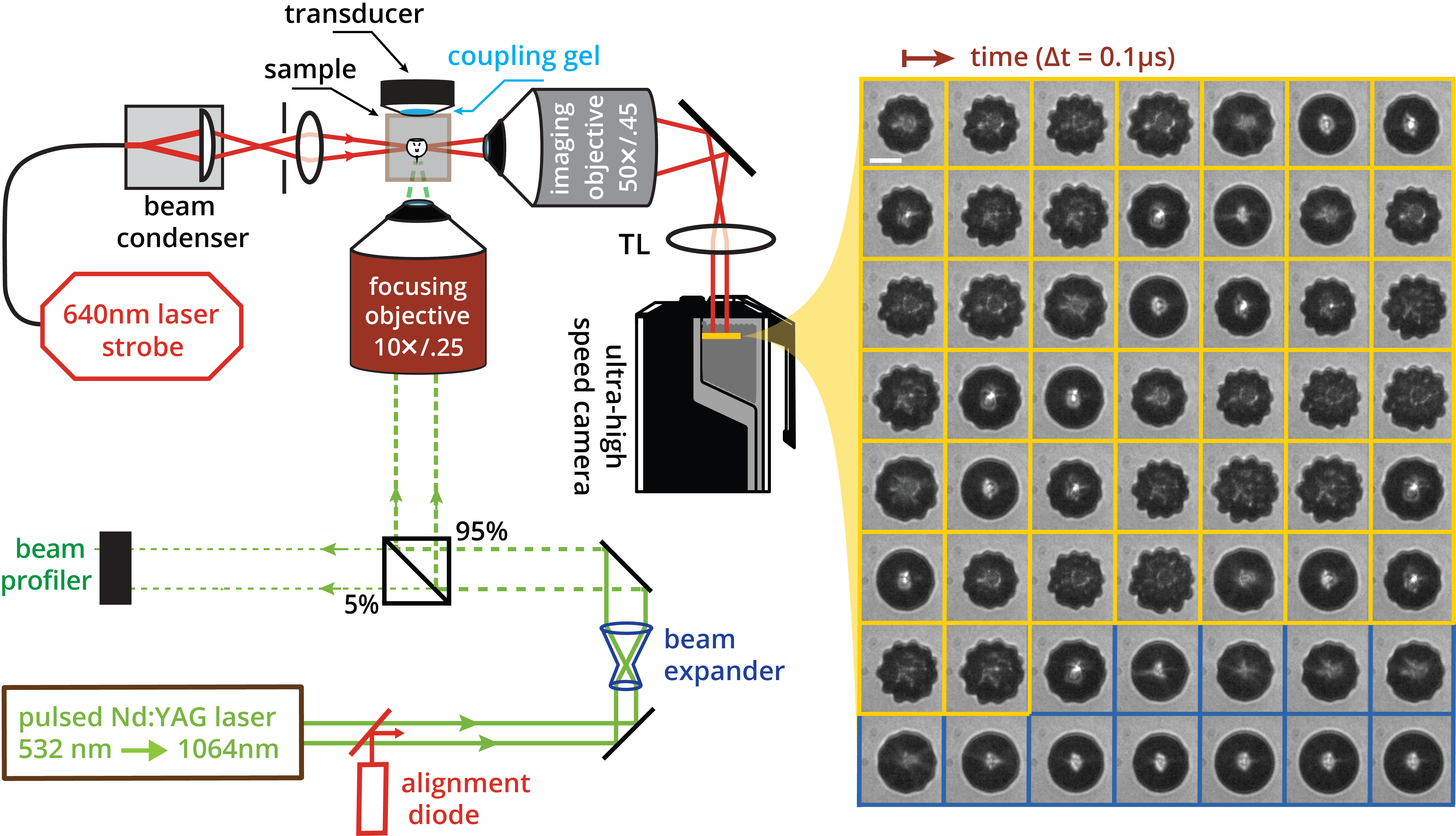}
    \caption{Experimental setup (drawn simplified) generates micro-cavitation using high energy laser, excites surface oscillations via in-house fabricated 10 mm diameter transducer, and records dynamics at \SI{E7}{} frames per second, \SI{1}{\milli\second} post cavitation. 
    The maize-outlined frames show when the driving \SI{750}{\kilo\hertz} ultrasound was active, while frames outlined in blue show inactive ultrasound. 
    Scale bar: \SI{20}{\micro\meter}.}
    \label{fig:Exp_Setup}
\end{figure*}

\Cref{fig:Exp_Setup} shows the experimental setup and follows our previously published work \cite{Abeid2024}.
The set-up applies to both experiments: (i) laser-induced inertial cavitation (LIC) then ultrasound forced and (ii) LIC with large amplitude radial oscillations.
For the latter experiments, an ultrasound transducer was not used. 
Images are processed using an edge detection-based MATLAB (MathWorks, MA, USA) processing routine.
A single \SI{4}{\nano\second} pulse of a user-adjustable, frequency-doubled Q-switched \SI{532}{\nano\meter} Nd:YAG laser (Continuum Minilite II, San Jose, CA) beam was aligned into a $10$X/0.25 NA high-power microspot focusing objective (LMH-10X-532; Thorlabs, Newton, NJ). 
The laser beam was directed using three broadband dielectric mirrors (BB1-E02; Thorlabs) and three short-pass dichroic mirrors. 
For alignment purposes, a dichroic mirror (DMSP605; Thorlabs) was aligned with a collimated continuous-wave laser diode (CPS635R; Thorlabs) to the pulsed beam to provide visible guidance. 
Prior to reaching the focusing objective, the collimated beam was expanded using a $2$X fixed magnification beam expander (GBE02-A; Thorlabs) to match the back aperture of the objective and reduce the risk of optical damage. 
Infrared wavelengths were removed by a second dichroic mirror (HBSY052; Thorlabs), which redirected unwanted spectral components into a beam dump (LB2; Thorlabs).

For the ultrasound forced experiments, \SI{1}{\milli\second} after cavitation was induced, small radial oscillations were achieved via an XY translation-stage-mounted, \SI{10}{\milli\meter} diameter cylindrical acoustic transducer.
The ultrasonic forcing excited the previously generated bubble surface oscillations with a $50$ cycle sinusoidal wave at \SI{750}{\kilo\hertz} and \SI{720}{\kilo\pascal}. 
An ultra-high-speed imaging camera (HPV-X2; Shimadzu, Kyoto, Japan) captured the bubble oscillations event at \SI{E7} frames per second. 

The LIC experiments with nonspherical bubbles are part of a larger dataset of previously published data in \citet{YANG2020}. 
Additional details on experimental setup differences and gel preparation are in \cite{YANG2020, supplementary_information}.

\subsection{Data processing}
A spectral interpolation approach is used to extract the perturbation amplitudes from the data points along the bubble surface~\cite{supplementary_information}.
A global search method (\texttt{bayesopt}) hybridized with a local search method (\texttt{lsqcurvefit}) in MATLAB is used to optimize the present model to the experimental data about the material properties $\mu, \, G, $ and $\alpha$.
For small radial oscillations, there is no effect on the dynamics from the strain stiffening parameter $\alpha$ and is therefore omitted from the optimization.
The surface tension is set to \SI{32}{\milli\newton\per\meter} \cite{Kikuchi_1991} and \SI{56}{\milli\newton\per\meter} \cite{Yang2021} for the gelatin gels and polyacrylamide, respectively.

\section{Results and discussion}
\label{sec:results}

\subsection{Small amplitude radial oscillations}
\begin{figure*}
    \centering
    \includegraphics[width=0.8\linewidth]{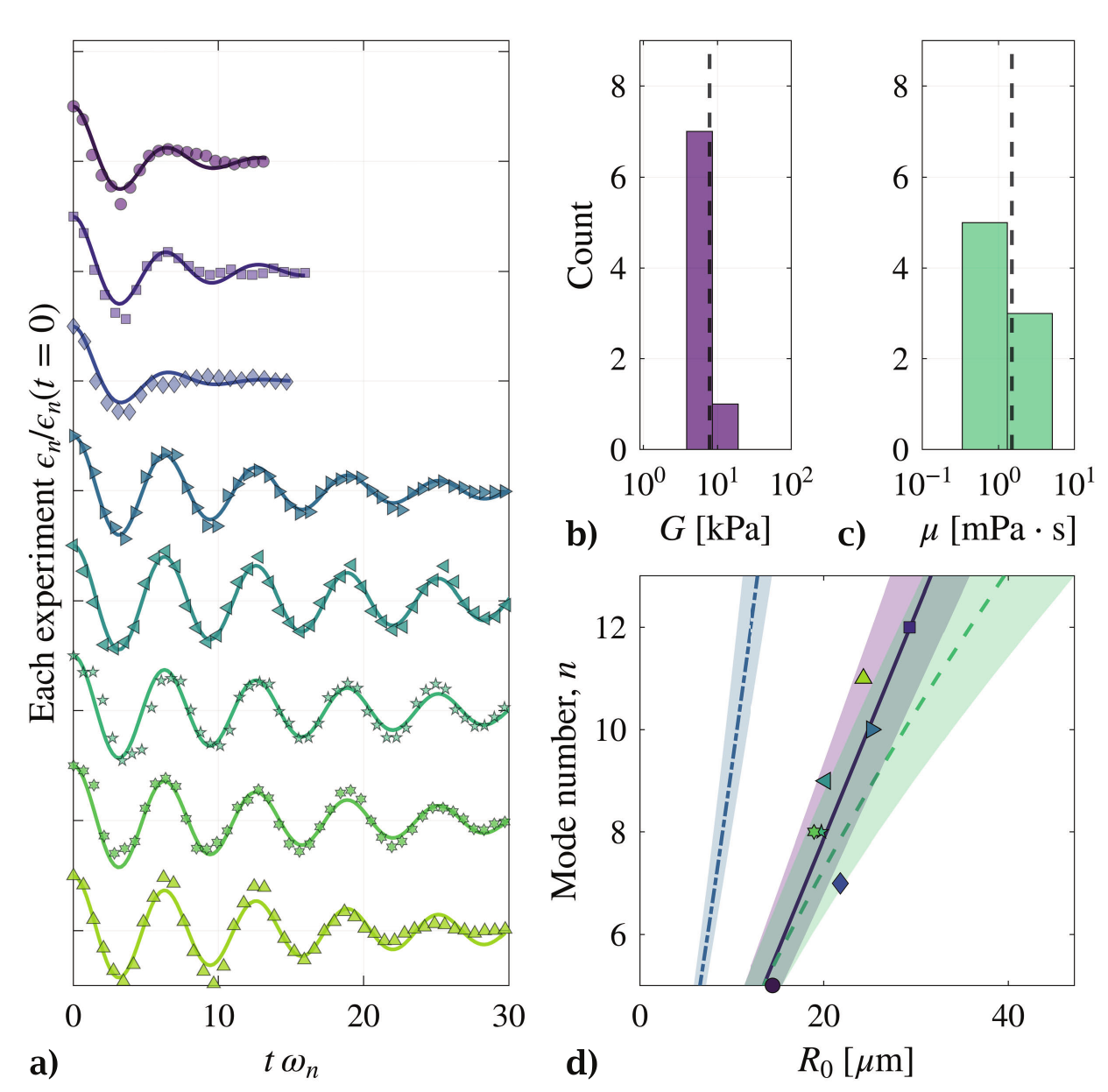}
    \caption{\textbf{a)} Evolution of the normalized amplitude for the most unstable mode for  10$\%$ gelatin, experimental data (symbols) and model prediction (lines), different colors correspond to different experiments. \textbf{b)} and \textbf{c)} Histograms of the calibrated shear modulus and viscosity, respectively. See \cite{supplementary_information} for all calibrated values. 
    \textbf{d)} Most unstable mode as a function of equilibrium radius. Symbols and purple shaded regions/lines: present experimental data and current theory, respectively; green: \citet{Murakami2020} model; and  steel blue: \citet{Yang2021} model. 
    Shaded regions around the mean predictions are from using the standard deviation of the distribution of calibrated shear moduli.}
    \label{fig:gel10_results}
\end{figure*}
\begin{figure}
    \centering
    \includegraphics[width=0.9\linewidth]{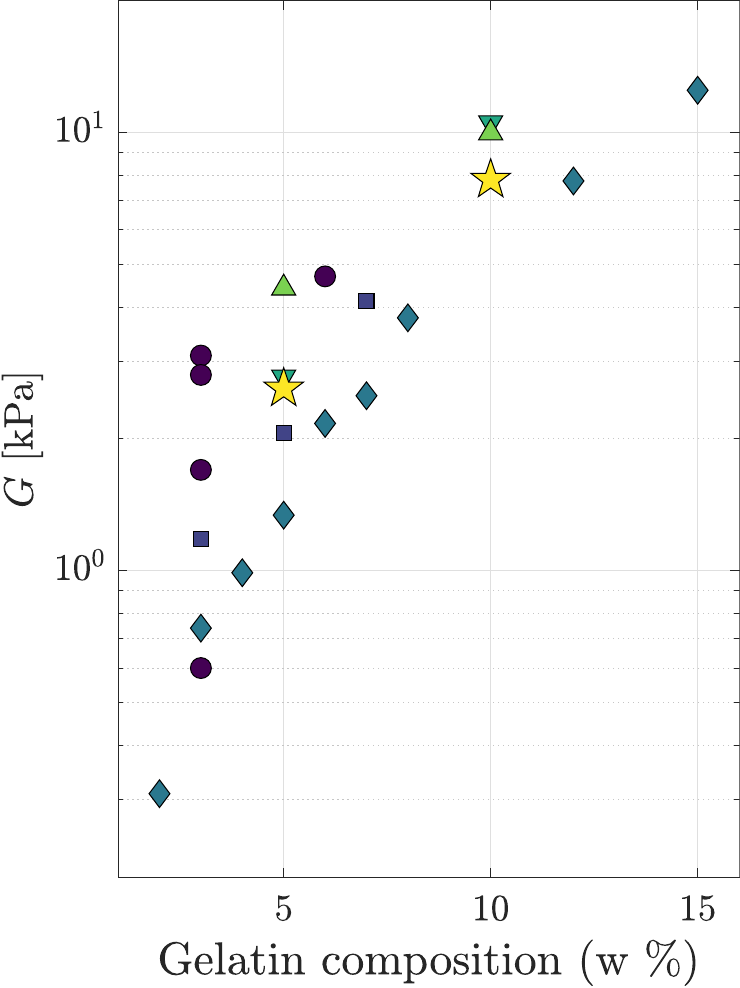}
    \caption{Shear modulus as a function of gelatin gel composition from the literature compared with current work. 
    Purple circles: \citet{Hamaguchi2015}; dark blue squares: \citet{Yoon2011}; steel blue diamonds: \citet{Bot1996}; dark green downward triangles: \citet{Abeid2024_droplet} quasi-static; light green upward triangles: \citet{Abeid2024_droplet} high strain rate; and yellow stars: current work.} 
    \label{fig:material_props_lit}
\end{figure}
\Cref{fig:gel10_results} shows a) the data and model predictions, b) and c) extracted parameters, and d) the most unstable mode as a function of equilibrium radius for the 10$\%$ gelatin gel.
$5\%$ gelatin gel results are in \cite{supplementary_information}.
Calibrated values for both viscosity and elasticity for each experiment are listed in \cite{supplementary_information}.
The mean values of the shear moduli calibrated to \cref{eq:pert_evo_linear} are $\overline{G}_{5\%} = 2.61$ $\SI{} {\kilo\pascal}$ and $\overline{G}_{10\%} =  7.81 $ $\SI{}{\kilo\pascal}$ for the 5$\%$ and 10$\%$ gelatin gels respectively.
\Cref{fig:material_props_lit} summarizes shear modulus across selected studies with various gelatin gel compositions.
The mean calibrated shear modulus and its scaling with composition is consistent between this study and previously measured values found in the literature. 
For both gelatin gels in this study, the calibrated viscosity is close to that of water, $\overline{\mu}_{5\%} = 1.6 $ \SI{}{\milli\pascal\cdot\second} and $\overline{\mu}_{10\%} = 1.5 $ \SI{}{\milli\pascal\cdot\second} which is an order of magnitude lower than the values found in the literature~\cite{supplementary_information}.

Aside from the high strain rate data of \cite{Abeid2024_droplet}, experiments from the literature were conducted at low (approximately quasi-static) strain rates.
The study by \citet{Abeid2024_droplet} showed that, using IMR, gelatin gels can become stiffer when subjected to high strain rates.
However, since the material shows a range of strain rates during microcavitation rheometry, the material properties from these experiments are effectively strain- and strain-rate-averaged.
For the current experiments, the strain rate can be calculated exactly using the rate of strain tensor, presented in \cite{supplementary_information}.
In \cref{fig:gel10_results}, $\dot{R} \approx 0$, and we assume that the perturbations are axisymmetric.
Under these assumptions the maximum component of the strain rate tensor for time and space ($r$, $\theta$, $\phi$) is, 
\begin{equation*}
    \max_{\psi, \dot{R} \approx 0}\left[\mathbf{D} \right] =  \left|D_{11}|_{r = \mathscr{R}(\theta=0, \phi)}\right| = (n+2)\sqrt{\frac{2n+1}{4\pi}}\dot{\epsilon}_n.
\end{equation*}
We then compute a time-averaged maximum strain rate, 
\begin{equation*}
    \frac{\overline{D}_{\text{max}}}{(n+2)}\sqrt{\frac{4\pi}{2n+1}} = \int_{t_i}^{t_f}\dot{\epsilon}_n dt 
    =\frac{\epsilon_n(t=0)}{\Delta t}, 
\end{equation*}
where $\Delta t$ is the experimental time window assuming that $\epsilon_n(\Delta t)\approx 0$.
We compute this value for each experiment and find an average of $\SI{0.17}{} \pm 0.1 \times 10^6$ and $\SI{0.19}{} \pm 0.07 \, \times 10^6 \, \SI{}{\second^{-1}}$ for the $5\%$ and $10\%$ gelatin gels, respectively.
The discrepancy of our calibrated viscosity values compared to the literature may be due to the sustained high strain rates of these experiments, suggesting that gelatin may exhibit a strain rate dependent viscosity.
%It has been previously suggested that materials subjected to LIC may be better described by non-Newtonaian models, \cite{Tzoumaka2023, Spratt2021}.

Lastly, we compare the most unstable mode predicted from the different models with the calibrated value for elasticity shown in \cref{fig:gel10_results} d).
We solve the parametric instability condition with the mean and $\pm \, \text{std}(G)$ to generate the predictive region for each model.
As expected, the experimental data lies directly within the region predicted by the present theory.
Additionally, the linear experimental data suggests better agreement with the present model than the model by \citet{Murakami2020}.

\subsection{Large amplitude radial oscillations}

\begin{figure*}
    \centering
    \includegraphics[width=0.95\linewidth]{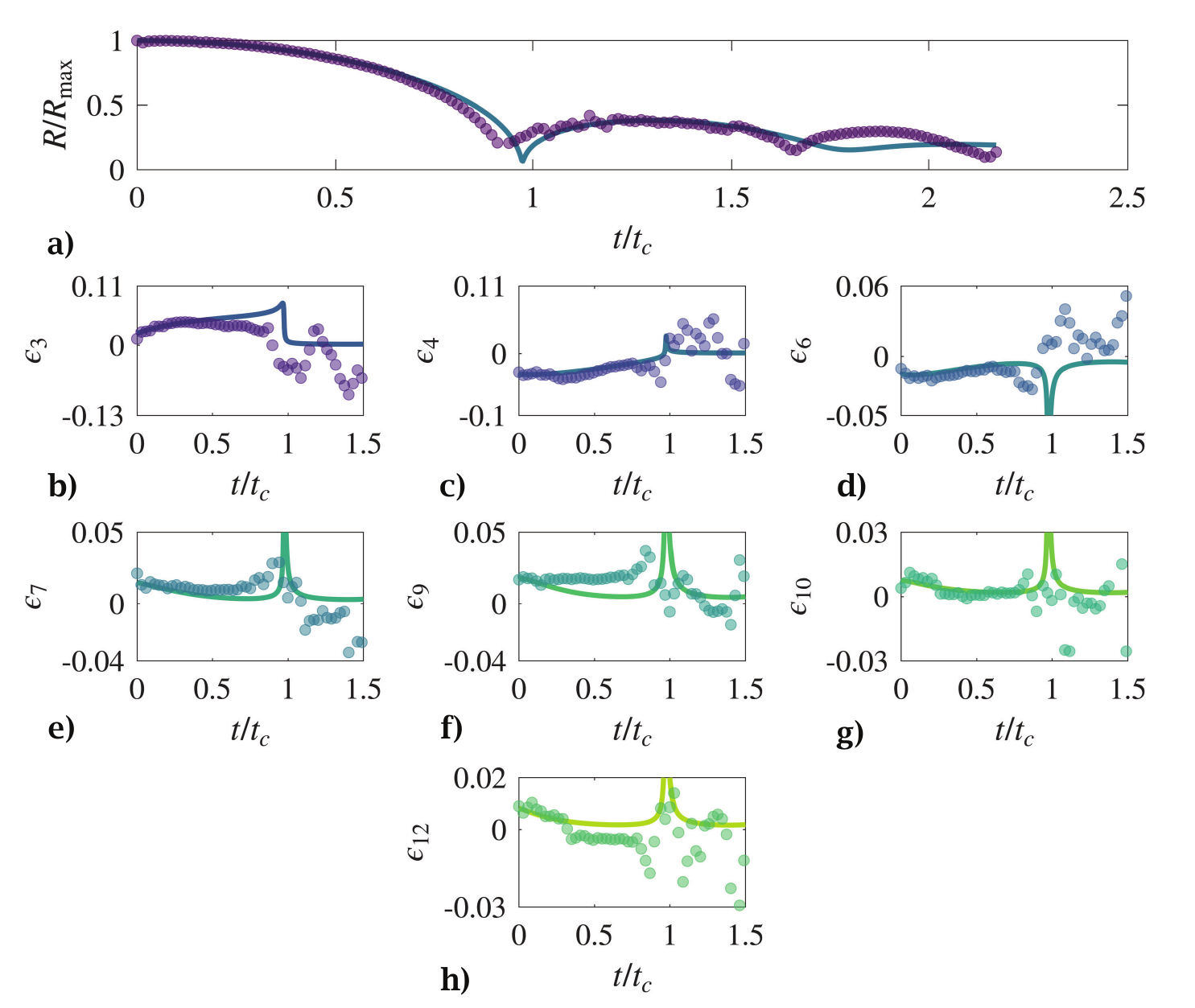}
    \caption{Non-spherical laser induced bubble \textbf{a}) mean radius evolution in time, \textbf{b-h})  third, fourth, sixth, seventh, ninth, tenth, and twelfth mode amplitude over time. 
    Circles: experimental data; and lines: calibrated model prediction.}
    \label{fig:LIC_exp3}
\end{figure*}
Four experimental trials are considered where the bubble surface exhibits small perturbations at its maximum radius~\cite{supplementary_information}.
\Cref{fig:LIC_exp3}a shows one set of the radial and surface perturbation modes experimental data and calibrated model prediction. 
Modes in the optimization and testing data for the calibrated model are shown in frames b-f and g-h, respectively.
The calibrated model agrees with experimental data before collapse.

Near the point of the collapse the perturbation prediction deviates from the experimental data.
There are three explanations.
First, although the perturbation amplitudes are exceedingly small, their coefficients $\eta$ and $\xi$ become large near collapse since they depend on $\dot{R}$ and $\ddot{R}$.
This leads to large $\dot{\epsilon}_n$ and $\ddot{\epsilon}_n$ near collapse.
Higher-order models capture interaction terms between different perturbation modes that are important when the magnitude of their time derivatives is large~\cite{Shaw2025JMF}.
Secondly, since the resolution of the camera is constant, as the bubble becomes smaller, the amount of data to construct the bubble surface becomes sparse.
Thirdly, through imaging, we only access a two dimensional projection of the non-spherical bubble surface and assume perturbations are within that plane.
Low variability of out of plane perturbations near the maximum radius (from small values of $\dot{R}$) may explain the agreement early in the experiment where the radial motion is approximately a linear deviation from the maximum radius.

Introducing the non-spherical perturbations into the inverse characterization procedure refines the calibrated material property values.
The present calibrated material parameters are $\overline{\alpha} = 0.083 \pm 0.048$ and $\overline{\mu} = 0.15 \pm 0.073$ \SI{}{\pascal\second}, see \cite{supplementary_information} for calibrated material properties for each dataset.
The mean viscosity is within 20$\%$ of the previously calibrated value of $\mu = 0.186 \pm 0.194$ \SI{}{\pascal\second} from \citet{YANG2020}.
On the other hand, the mean strain hardening parameter is over $80\%$ different from the previously calibrated value of $\alpha = 0.48 \pm 0.14$. 
The discrepancy is due to calibration being limited to first collapse in the study by \citet{YANG2020}.
%However, constraining the calibration problem by increasing the number of independent observables yields refined $\alpha$ and $\mu$ values.

\section{Conclusions}
\label{sec:conclusions}
We present a model for the evolution of bubble surface instabilities within a strain stiffening viscoelastic material.
In the linear radial oscillation regime and with zero shear modulus, our model agrees with the irrotational model from \cite{Prosperetti1977}.
We compare our model to two sets of experiments, small radial oscillations from ultrasound forcing within $5\%$ and $10\%$ gelatin solutions and large radial oscillations from LIC in polyacrilamide.
The experimental data shows a linear relationship between the most unstable mode and equilibrium radius in viscoelastic materials which is only predicted by the current theory.
The LIC bubble experiments showed small surface perturbations at the maximum radius which grow throughout the collapse phase.
Despite having a two-dimensional view of the bubble and modeling the perturbations in a linear fashion, agreement between the experimental data and the model is observed.
We validated our model for different materials and dynamical regimes from small radial oscillations to large, inertial collapses with non-spherical perturbations.
The modeling approach taken in this study can easily be applied to other material models by considering a different Cauchy stress tensor.
Moreover, a single simulation of the presented current theory for the case of large radial oscillation runs on the order of about 1 minute locally (Dell Alienware x15 R2 with intel CORE-i9).
In contrast, a full three-dimensional numerical simulation of the problem can exceed run times of 48 hours on 48 CPU cores (GCS Supercomputer SuperMUC-NG at Leibniz Supercomputing Centre) \cite{ALPACASim}.
Future studies will investigate the use of non-spherical ultrasound forced bubbles for characterization across a range of strain rates by modulating the forcing frequency and bubble size.

\section{Acknowledgments}
MRJ acknowledges support from the U.S. Department of Defense under the DEPSCoR program Award No. FA9550-23-1-0485 (PM Dr.~Timothy~Bentley).
JBE, MRJ, and JY gratefully acknowledge support from the U.S. National Science Foundation (NSF) under Grant Nos. 2232426, 2232427, and 2232428, respectively.
Funding agencies were not involved in study design; in the collection, analysis and interpretation of data; in the writing of the report; or in the decision to submit the article for publication.
The opinions, findings, and conclusions, or recommendations expressed are those of the authors and do not necessarily reflect the views of the funding agencies.

\printcredits

%% Loading bibliography style file
%\bibliographystyle{model1-num-names}
\bibliographystyle{cas-model2-names}

% Loading bibliography database
\bibliography{refs}

\end{document}

\let\WriteBookmarks\relax
\def\floatpagepagefraction{1}
\def\textpagefraction{.001}
\shorttitle{Microbubble surface instabilities in a strain stiffening viscoelastic material}
\shortauthors{S. Remillard et~al.}

\title [mode = title]{Microbubble surface instabilities in a strain stiffening viscoelastic material}                      
% \tnotemark[1,2]

%\tnotetext[1]{This document is the results of the research
 %  project funded by the National Science Foundation.}

%\tnotetext[2]{The second title footnote which is a longer text matter
 %  to fill through the whole text width and overflow into
 %  another line in the footnotes area of the first page.}

\author[1]{Sawyer Remillard}[orcid=0000-0002-5469-6168]
\cormark[1]
\credit{Conceptualization of this study, Methodology, Software, Data curation, Writing - Original draft preparation}
\author[2]{Bachir A. Abeid}[orcid=0009-0009-0859-8434]
\credit{Methodology, Data curation, Writing - review \& editing}
\author[3]{Timothy L. Hall}[]
\credit{Conceptualization of this study, Writing - review \& editing}
\author[3]{Jonathan R. Sukovich}[]
\credit{Conceptualization of this study, Writing - review \& editing}
\author[4]{Jacob Baker}[orcid=]
\credit{Methodology, Data curation, Writing - review \& editing}
\author[5,6]{Jin Yang}[orcid=0000-0002-5967-980X]
\credit{Conceptualization of this study, Data curation, Writing - review \& editing, Funding Acquisition}
\author[2]{Jonathan B. Estrada}[orcid=0000-0003-1083-4597]
\credit{Conceptualization of this study, Writing - review \& editing, Funding Acquisition}
\author[1]{Mauro Rodriguez Jr.}[orcid=0000-0003-0545-0265]
\cormark[2]
\credit{Conceptualization of this study, Writing - original draft, Writing - review \& editing, Funding Acquisition, Project Administration, Supervision}

\affiliation[1]{organization={School of Engineering, Brown University}, 
                city={Providence},
                postcode={02912}, 
                state={RI},
                country={USA}}

\affiliation[2]{organization={Department of Mechanical Engineering, University of Michigan},
                city={Ann Arbor},
                postcode={48109}, 
                state={MI},
                country={USA}}

\affiliation[3]{organization={Department of Biomedical Engineering, University of Michigan},
                city={Ann Arbor},
                postcode={48109}, 
                state={MI},
                country={USA}}

\affiliation[4]{organization={Department of Biomedical Engineering, The University of Texas at Austin},
                city={Austin},
                postcode={78712}, 
                state={TX},
                country={USA}}
\affiliation[5]{organization={Department of Aerospace Engineering and Engineering Mechanics, The University of Texas at Austin},
                city={Austin},
                postcode={78712}, 
                state={TX},
                country={USA}}
\affiliation[6]{organization={Texas Materials Institute, The University of Texas at Austin},
                city={Austin},
                postcode={78712}, 
                state={TX},
                country={USA}}

\cortext[cor1]{Corresponding author}
\cortext[cor2]{Principal corresponding author}

\section{Kinematically-consistent analytical model}
\label{sec:theoretical_model}
The linearized deformation gradient tensor is
\begin{equation}
 \begin{aligned}
&\boldsymbol{F}=\frac{\partial \boldsymbol{x}}{\partial \boldsymbol{x}_o} =
 \left[\begin{array}{ccc}
 \frac{\partial r}{\partial r_o} & \frac{1}{r_o}\frac{\partial r}{\partial \theta_o} & \frac{1}{r_o \sin\theta_o}\frac{\partial r}{\partial \phi_o} \\
 r\frac{\partial \theta}{\partial r_o} & \frac{r}{r_o}\frac{\partial \theta}{\partial \theta_o} & \frac{r}{r_o \sin\theta_o}\frac{\partial \theta}{\partial \phi_o} \\
 r \sin\theta\frac{\partial \phi}{\partial r_o} & \frac{r \sin\theta}{r_o}\frac{\partial \phi}{\partial \theta_o} & \frac{r \sin\theta}{r_o\sin\theta_o} \frac{\partial \phi}{\partial \phi_o}
\end{array}\right] = \\
& \left[\begin{array}{ccc}
\frac{\partial \left( r_s+\epsilon_nr_1\right)}{\partial r_o} &  \frac{\epsilon_n }{r_o}\frac{\partial r_1}{\partial \theta_o} & \frac{\epsilon_n }{r_o \sin\theta_o} \frac{\partial r_1}{\partial \phi_o} \\
\epsilon_n r_s\frac{\partial \theta_1}{\partial r_o}    & \frac{r_s}{r_o}+\epsilon_n \left(\frac{r_1}{r_o} +  \frac{r_s}{r_o} \frac{\partial \theta_1}{\partial \theta_o}\right)   & \epsilon_n \frac{r_s}{r_o \sin \theta_o}\frac{\partial \theta_1}{\partial \phi_o} \\
\epsilon_n r_s\sin\theta_o\frac{\partial \phi_1}{\partial r_o}  & \epsilon_n \sin \theta_o \frac{r_s}{r_o}\frac{\partial \phi_1}{\partial \theta_o} & \frac{r_s}{r_o}+ \epsilon_n \left(\frac{r_1}{r_o} + \frac{r_s}{r_o} \left(\theta_1 \cot\theta_o + \frac{\partial \phi_1}{\partial \phi_o}\right) \right)
\end{array}\right].
\label{eq:full_def_grad}
 \end{aligned}
\end{equation}

The linearized deformation and inverse deformation maps are
\begin{subequations}
\begin{equation}
    r(\psi_o, t) = r_s + \epsilon_n \, \frac{R^{n+3}}{r_s^{n+2}}Y_n^m(\theta_o, \phi_o), 
\end{equation}
\begin{equation}
    \theta(\psi_o, t) =\theta_o- \epsilon_n \, \frac{R^{n+3}}{(n+1)r_s^{n+3}}\frac{\partial}{\partial\theta_o}Y_n^m(\theta_o, \phi_o), 
\end{equation}
\begin{equation}
    \phi(\psi_o, t) = \phi_o - \epsilon_n \, \frac{R^{n+3}}{(n+1)r_s^{n+3}}\frac{1}{\sin^2\theta_o}\frac{\partial}{\partial\phi_o}Y_n^m(\theta_o, \phi_o).
\end{equation}    
\end{subequations}
\begin{subequations}
\begin{equation}
    r_o(\psi, t) = \varrho - \epsilon_n \, \frac{R^{n+3}}{r^n\varrho^{2}}\YY,
\end{equation}
\begin{equation}
    \theta_o(\psi, t) =\theta + \epsilon_n \, \frac{R^{n+3}}{(n+1)r^{n+3}}\frac{\partial}{\partial\theta}\YY,
\end{equation}
\begin{equation}
    \phi_o(\psi, t) = \phi + \epsilon_n \, \frac{R^{n+3}}{(n+1)r^{n+3}}\frac{1}{\sin^2\theta}\frac{\partial }{\partial\phi}\YY, 
\end{equation}   
\label{ap:def_maps}
\end{subequations}
respectively, where $\varrho =\left(r^3-R^3+R_o^3 \right)^{1/3} $.
The linearized Eulerian velocity and acceleration are 
\begin{subequations}
\begin{equation}
        v_r = \frac{R^2 \dot{R}}{r^2}+\frac{R^{n+2}}{r^{n+5}}\left(\epsilon_n  \dot{R} \left((n+3) r^3-n R^3\right)+r^3 R \dot{\epsilon}_n\right)\YY,
\end{equation}
\begin{equation}
        v_{\theta} =-\frac{R^{n+2}}{(n+1)r^{n+5}} \left(\epsilon_n \dot{R}(n+3) \left(r^3-R^3\right) +r^3 R \dot{\epsilon}_n\right)\frac{\partial \YY}{\partial \theta},
\end{equation}
\begin{equation}
        v_{\phi}  = \frac{R^{n+2}}{\sin \theta (n+1)r^{n+5}}\left(\epsilon_n \dot{R}(n+3) \left(r^3-R^3\right) +r^3 R \dot{\epsilon}_n\right)\frac{\partial \YY}{\partial \phi},
\end{equation}
\label{eq:velocity}
\end{subequations}
\begin{subequations}
\begin{equation}
\begin{aligned}
    a_r &= \frac{2R\dot{R}^2+R^2\ddot{R}}{r^2}
    - \frac{2R^4\dot{R}^2}{r^5}  + \frac{R^{n+1}}{r^{n+8}}\bigg[r^3 R 
    \left(\epsilon_n\left((n+3) r^3-n R^3\right)\ddot{R}
    + r^3 R \ddot{\epsilon}_n\right) \\
    & + 2 r^3 R \dot{R}\dot{\epsilon}_n 
    \Big((n+3) r^3-(n+2) R^3\Big) + \epsilon_n \dot{R}^2
    \left((n+2)(n+3) r^6 \right. \left.- 2 (n(n+6)+6) r^3 R^3 
    + n (n+7) R^6\right) \bigg] \YY,
\end{aligned}
\end{equation}
\begin{equation}
    \begin{aligned}
        a_{\theta}& = -\frac{R^{n+1}}{(n+1)r^{n+8}} \Bigg(\epsilon_n  (n+3) \left(r^3-R^3\right) \left(\left((n+2) r^3-(n+4) R^3\right) \dot{R}^2+r^3 R \ddot{R}\right)\\
        & + r^3 R \left(2 \left((n+3) r^3-(n+2) R^3\right) \dot{R} \dot{\epsilon}_n +r^3 R \ddot{\epsilon}_n\right) \Bigg) \frac{\partial \YY}{\partial \theta},
    \end{aligned}
\end{equation}
\begin{equation}
    \begin{aligned}
        a_{\phi} & = -\frac{R^{n+1}}{\sin\theta (n+1)r^{n+8}} \Bigg(\epsilon_n  (n+3) \left(r^3-R^3\right) \left(\left((n+2) r^3-(n+4) R^3\right) \dot{R}^2+r^3 R \ddot{R}\right)\\
        &  + r^3 R \left(2 \left((n+3) r^3-(n+2) R^3\right) \dot{R} \dot{\epsilon}_n +r^3 R \ddot{\epsilon}_n\right) \Bigg) \frac{\partial \YY}{\partial \phi}.
    \end{aligned}
\end{equation}  
\label{eq:acceleration}
\end{subequations}

\subsection{Linearized quantities for momentum balance}
\label{appendix:linear}

The left Cauchy-Green deformation tensor given the linearized deformation is
\label{ap:lin_tensors_mom_bal}
\begin{equation}
    \boldsymbol{B} = 
    \begin{bmatrix}
        B_{11} & B_{12} & B_{13}\\
        B_{12} & B_{22} & B_{23}\\
        B_{13} & B_{23} & B_{33}
    \end{bmatrix},
\end{equation}
where 
\begin{subequations}\label{eq:B-block}
\begin{equation}
    B_{11} = \frac{\varrho^4}{r^4} - 2\epsilon_n \varrho\,
         \frac{R^{n+3}\Big((n+2)r^3 + n(R_o^3 - R^3)\Big)\,}
              {r^{n+7}}\,\YY,
\end{equation}
\begin{equation}
    B_{12} = \epsilon_n\,\frac{R^{n+3}}{r^{n+7}}
          \left(
            \frac{(n+3)\,\varrho^4}{n+1}
            + \frac{r^6}{\varrho^2}
          \right) \frac{\partial \YY}{\partial \theta},
\end{equation}
\begin{equation}
    B_{13} = \epsilon_n\,\frac{R^{n+3}}{r^{n+7}}
          \left(
            \frac{(n+3)\,\varrho^4}{n+1}
            + \frac{r^6}{\varrho^2}
          \right) \csc\theta\, \frac{\partial \YY}{\partial \phi},
\end{equation}
\begin{equation}
    B_{22} = \frac{r^2}{\varrho^2}  + 
        2\epsilon_n\,\frac{R^{n+3}}{r^{n+1}}
         \left(
           \frac{r^3}{\varrho^5}\,\YY
           - \frac{1}{(n+1)\,\varrho^2}\,
             \frac{\partial^2 \YY}{\partial {\theta}^2}
         \right),
\end{equation}
\begin{equation}
    B_{23} = \frac{2\epsilon_n\,R^{n+3}}{(n+1)\,r^{n+1}\,\varrho^2}
         \,\csc\theta \left(
           \cot\theta\,\frac{\partial \YY}{\partial \phi}
           - \frac{\partial^2 \YY}{\partial \theta \partial \phi}
         \right),
\end{equation}
\begin{equation}
    \begin{split}
    B_{33} &= \frac{r^2}{\varrho^2} + 2\epsilon_n\,\frac{R^{n+3}}{r^{n+1}}
         \left(
           \frac{(n+1)r^3 + n(R_o^3-R^3)}{\varrho^5}\,\YY \right. \left. + \frac{1}{(n+1)\,\varrho^2}\,\frac{\partial^2 \YY}{\partial \theta^2}
         \right).
    \end{split}
\end{equation}
\end{subequations}
Using the Eulerian velocity, we can take the spatial gradient and average it with its transpose to obtain the rate of strain tensor,
\begin{equation}
    \boldsymbol{D} = 
    \begin{bmatrix}
        D_{11} & D_{12} & D_{13}\\
        D_{12} & D_{22} & D_{23}\\
        D_{13} & D_{23} & D_{33}
    \end{bmatrix},
\end{equation}
where
\begin{subequations}
\begin{equation}
\begin{split}
D_{11} &= -\frac{2 R^2 \dot R}{r^3} - \frac{R^{n+2}}{r^{n+6}} \left(
  \left( (n+2)(n+3) r^3 - n(n+5) R^3\right)\,\epsilon_n \dot R
  \right.  + \left. (n+2) r^3 R\,\dot{\epsilon}_n
\right)\,\YY,
\end{split}
\end{equation}
\begin{equation}
\begin{split}
D_{12} &= \frac{R^{n+2}}{2 (n+1) r^{n+6}}\left(
  \big((n+2)(n+3) r^3 - (n(n+5)+9) R^3\big)\,\epsilon_n \dot R \right.+ \left. (n+2) r^3 R\,\dot{\epsilon}_n
\right)\,\frac{\partial \YY}{\partial \theta},
\end{split}
\end{equation}
\begin{equation}
    \begin{split}
D_{13} &= \frac{R^{n+2}}{2 (n+1) r^{n+6}}\left(
  \big((n+2)(n+3) r^3 \right.  \left. - (n(n+5)+9) R^3\big)\,\epsilon_n \dot R
  + (n+2) r^3 R\,\dot{\epsilon}_n
\right)\,\csc\theta\,\frac{\partial \YY}{\partial \phi},
    \end{split}
\end{equation}
\begin{equation}
\begin{split}
D_{22} &= \frac{R^2 \dot R}{r^3} + \frac{R^{n+2}}{r^{n+6}}\Big[
  \big((n+3) r^3 - n R^3\big)\,\epsilon_n \dot R
  + r^3 R\,\dot{\epsilon}_n
    \Big]\,\YY - \frac{1}{n+1}\Big(
  (n+3)\epsilon_n (r^3 - R^3)\dot R
  + r^3 R\,\dot{\epsilon}_n
\Big)\,\frac{\partial^2 \YY}{\partial \theta^2},
\end{split}
\end{equation}
\begin{equation}
\begin{split}
D_{23} &= \frac{R^{n+2}}{(n+1)\,r^{n+6}}\,
\csc\theta\Big(
  (n+3)\epsilon_n (r^3 - R^3)\dot R
  + r^3 R\,\dot{\epsilon}_n
\Big) \Big(
  \cot\theta\,\frac{\partial \YY}{\partial \phi}
  - \frac{\partial^2 \YY}{\partial \theta \partial \phi}
\Big),
\end{split}
\end{equation}
\begin{equation}
\begin{split}
D_{33} &= \frac{R^2 \dot R}{r^3} + \frac{R^{n+2}}{r^{n+6}}\left(
  \epsilon_n \dot R\big((n+1)(n+3) r^3 \right.\left.- n(n+4) R^3\big)
  + (n+1) r^3 R\,\dot{\epsilon}_n
\right)\,\YY \\
  &+ \frac{1}{n+1}\Big(
  (n+3)\epsilon_n (r^3 - R^3)\dot R
  + r^3 R\,\dot{\epsilon}_n
\Big)\,\frac{\partial^2 \YY}{\partial \theta^2}.
\end{split}
\end{equation}
\end{subequations}

For the calculation of the strain rates that the material undergoes during the experiment, we evaluate the rate of strain tensor at the bubble wall,
\begin{equation}
    \boldsymbol{D}|_{r = \mathscr{R}(\theta, \phi)}= 
    \begin{bmatrix}
        D_{11} & D_{12} & D_{13}\\
        D_{12} & D_{22} & D_{23}\\
        D_{13} & D_{23} & D_{33}
    \end{bmatrix}_{r = \mathscr{R}(\theta, \phi)},
\end{equation}
where
\begin{subequations}
    \begin{equation}
        D_{11}|_{r = \mathscr{R}(\theta, \phi)} = -\frac{2 \dot{R}}{R}-(n+2) \dot{\epsilon}_n \YY,
    \end{equation}
    \begin{equation}
        D_{12}|_{r = \mathscr{R}(\theta, \phi)} = \frac{ -3 \epsilon_n \dot{R}+(n+2) R \dot{\epsilon}_n}{(n+1) R}\frac{\partial \YY}{\partial \theta} ,
    \end{equation}
    \begin{equation}
        D_{13}|_{r = \mathscr{R}(\theta, \phi)} = \frac{-3 \epsilon_n \dot{R}+(n+2) R \dot{\epsilon}_n}{(n+1) R}\frac{1}{\sin \theta}  \frac{\partial \YY}{\partial \phi} ,
    \end{equation}
    \begin{equation}
    \begin{aligned}
        D_{22}|_{r = \mathscr{R}(\theta, \phi)} & = \frac{\dot{R}}{R} + \dot{\epsilon}_n\left(\YY-  \frac{1}{n+1}\frac{\partial^2 \YY}{\partial \theta^2} \right) ,
        \end{aligned}
    \end{equation}
    \begin{equation}
       D_{23}|_{r = \mathscr{R}(\theta, \phi)} = \frac{ \dot{\epsilon}_n}{(n+1) }\frac{1}{\sin \theta} \left(\cot \theta \frac{\partial \YY}{\partial \phi} -\frac{\partial^2 \YY}{\partial \phi \, \partial \theta} \right) ,
    \end{equation}
    \begin{equation}
    \begin{aligned}
       D_{33}|_{r = \mathscr{R}(\theta, \phi)} &=  \frac{\dot{R}}{R} + \frac{\dot{\epsilon}_n}{(n+1)}\left((n+1)^2 \, \YY-  \frac{\partial^2 \YY}{\partial \theta^2} \right) ,
       \end{aligned}
    \end{equation}
    \label{eq:rate_of_strain_e}
\end{subequations}

\subsection{Coefficients for perturbation evolution equation}
\label{ap:pert_evo_append}
The damping coefficient and squared natural frequency for the non-linear perturbation evolution equation are
\begin{equation}
    \eta = \frac{4\left(5+2n \right)}{(4+n)}\frac{\dot{R}}{R}+\frac{2 (n+1)(n+2)}{\rho R^2}\mu,
\end{equation}
\begin{equation}
\begin{aligned}
        \xi =&6\frac{n+2}{n+4} \frac{\dot{R}^2}{R^2}- \frac{n^2-n-8}{n+4}\frac{ \ddot{R}}{R}+\frac{ n(n+1)(7n+20)\dot{R}}{(n+6)\rho R^3}\mu+\frac{(n+1)(n-1)(n+2)\gamma}{\rho R^3} + \xi_{\text{nH}}G + \xi_{\alpha}\alpha G,
        \end{aligned}
\end{equation}
where
\begin{equation}
  \begin{split}
    \xi_{\text{nH}} &= \frac{n+1}{\lambda^{11}(n+6)(n+9)(n+12)\,\rho\,R_o^2} \Bigg[ 2\lambda^5\,(n^3+27n^2+234n+648)\,(-\lambda^6+2\lambda^3+n+1) \\
    &+ \lambda^3(\lambda^3-1)(n+13)\Big(108\lambda^6
       + (3\lambda^3-1)n^3
       + (2\lambda^6+35\lambda^3-9)n^2  + 2(15\lambda^6+43\lambda^3-7)n \Big)\,
     \\
     & \times{}_2F_1\!\left[-\tfrac{1}{3},\tfrac{n}{3}+4;\tfrac{n}{3}+5;1-\tfrac{1}{\lambda^3}\right] + (\lambda^3-1)\Big(-108\lambda^9
       + (\lambda^3-1)^2n^4
       + (16\lambda^6-37\lambda^3+19)n^3 \\
    & + (-30\lambda^9+54\lambda^6-266\lambda^3+140)n+ (-2\lambda^9+69\lambda^6-199\lambda^3+104)+ n^2 \Big)\,
      {}_2F_1\!\left[\tfrac{2}{3},\tfrac{n}{3}+4;\tfrac{n}{3}+5;1-\tfrac{1}{\lambda^3}\right]
    \Bigg],
  \end{split}
\end{equation}
and 
\begin{equation}
    \begin{split}
     \xi_{\alpha} & =  \frac{n+1}{\lambda ^{10} \rho  R_o}\Bigg[2 \left(2 \lambda ^6-3 \lambda ^4+2 \lambda ^3+3\right) \left(\lambda ^3-1\right)^2+2 \left(\lambda ^6-3 \lambda ^4+2\right) \left(2 \lambda ^3+n\right) + \frac{\lambda ^2}{(n+6) (n+9) (n+12)} \left(\frac{1}{\lambda ^3}-1\right)\\
     & \times \Bigg[3 \left(-2 \lambda ^9 (n+6) (n+9)+\lambda ^6 n (n+1) (n+6) (n+9)-\lambda ^3 n (n+2) (n+7) (2 n+19)+n (n+2) (n+7) (n+10)\right) \\
     &\times \,   _2F_1\left[\frac{2}{3},\frac{n}{3}+4;\frac{n}{3}+5;1-\frac{1}{\lambda ^3}\right]-\lambda ^3 \Bigg(3 (n+13) \left(-2 \lambda ^6 (n+6) (n+9)-\lambda ^3 n (n (3 n+35)+86)+n (n+2) (n+7)\right)\\
     & \times \, _2F_1\left[-\frac{1}{3},\frac{n}{3}+4;\frac{n}{3}+5;1-\frac{1}{\lambda ^3}\right]+(n+14) \Big(4 \lambda ^9 (n+6) (n+9)+2 \lambda ^6 (n+1) (n+2) (n+6) (n+9)\\
     & -\lambda ^3 (n (n (n (4 n+59)+229)+120)-324)+n (n (n (2 n+33)+175)+318)\Big) \, _2F_1\left[-\frac{2}{3},\frac{n}{3}+4;\frac{n}{3}+5;1-\frac{1}{\lambda ^3}\right]\Bigg)\\
     &+2 \Big(4 \lambda ^{12} (n+6) (n+9)+2 \lambda ^9 (n+1) (n+2) (n+6) (n+9)+\lambda ^6 (n+1) (n+6) (n+9) (n (n+6)+6)\\
     &-\lambda ^3 n (n (n (2 n (n+25)+427)+1411)+1398)+n (n+2) (n+11) (n (n+13)+39)\Big) \, _2F_1\left[\frac{1}{3},\frac{n}{3}+4;\frac{n}{3}+5;1-\frac{1}{\lambda ^3}\right] \Bigg] \Bigg],
    \end{split}
\end{equation}
where $_2F_1\left[\, \cdot\,\right]$ is the gaussian hypergeometric function.

\subsection{Coefficients for perturbation evolution equation, Yang et. al.}
\label{ap:Yang_coeff}
The work by both \citet{Yang2021} and \citet{Murakami2020} is formulated in terms of dimensional perturbation amplitude. 
For consistency, we consider the dimensional amplitude $a_n = \epsilon_n \, R$. 
The \citet{Yang2021} model coefficients are
\begin{equation}
    \eta = 5\frac{\dot{R}}{R}+4\frac{\mu}{\rho R^2}+\frac{n(n+1)\mu}{3\rho R^2},
\end{equation}
\begin{equation}
\begin{aligned}
    \xi &= 2\frac{\ddot{R}}{R} + 3\frac{\dot{R}^2}{R^2}+\frac{n(n+1)\dot{R}\mu }{\rho R^3}   + \frac{(n+2)(n-1) \gamma}{\rho R^3} + \frac{G}{\rho R_o^2}\left(\frac{n(n+1)}{\lambda^2+\lambda+1} +\frac{2(1+\lambda^3)}{\lambda^6}\right) + \frac{G\alpha(\lambda-1)^2}{5\rho R_o^2(\lambda^2+\lambda+1)\lambda^{10}} \\
    &\times \Bigg[10(n(n+1)+2)\lambda^{9}+6(n(n+1)+10)\lambda^{8}+3(n(n+1)+30)\lambda^{7} + (n(n+1)+110)\lambda^{6} + 120\lambda^{5}+ 120\lambda^{4}\\
    &+100\lambda^{3}+60\lambda^{2}+30\lambda+10\Bigg].
\end{aligned}
\end{equation}
After linearization, 
\begin{equation}
    \begin{aligned}
        \eta_{\rm l} =\frac{  \left(n(n+1)+12\right)\mu}{3 \rho  R_o^2}, \quad \quad \xi_{\rm l} = \omega_n^2 =  \frac{  \left(n(n+1)+12\right)G}{3 \rho  R_o^2} + \frac{(n+2)(n-1)\gamma}{\rho R_o^3}.
    \end{aligned}
    \label{eq:Yang_lin_coef}
\end{equation}

\subsection{Coefficients for perturbation evolution equation, Murakami et. al.}
\label{ap:Murakami_coeff}
The work done by \citet{Murakami2020} does not include the strain stiffening analysis. 
However, the deformation mapping for the computation of the Cauchy-Stress tensor matches the deformation mapping used in \citet{Yang2021}. 
We incorporate these terms into the theory developed by \citet{Murakami2020} to obtain the following coefficients, 
\begin{equation}
    \eta = 5 \frac{\dot{R}}{R}+2\frac{(n+1)(n+2)\mu}{\rho R^2}
\end{equation}
\begin{equation}
\begin{aligned}
    \xi &= (2-n)\frac{\ddot{R}}{R} + 3\frac{\dot{R}^2}{R} + 6\frac{n(n+1)\dot{R}\mu}{\rho R^3}  + \frac{(n+2)(n+1)(n-1) \gamma}{\rho R^3} + \frac{G(n+1)}{\rho R_o^2}\left(\frac{n(n+1)}{\lambda^2+\lambda+1} +\frac{2(1+\lambda^3)}{\lambda^6}\right)   \\
    &+\frac{G\alpha(n+1)(\lambda-1)^2}{5\rho R_o^2(\lambda^2+\lambda+1)\lambda^{10}} \Bigg[10(n(n+1)+2)\lambda^{9} +6(n(n+1)+10)\lambda^{8}+3(n(n+1)+30)\lambda^{7}+ (n(n+1)+110)\lambda^{6}\\
    &+ 120\lambda^{5}+ 120\lambda^{4}+100\lambda^{3}+60\lambda^{2}+30\lambda+10\Bigg],
\end{aligned}
\end{equation}
After linearization, 
\begin{equation}
    \begin{aligned}
         \eta_{\rm l} =\frac{2  (n+2)(n+1)\mu}{\rho  R_o^2}, \quad \quad \xi_{\rm l} = \omega_n^2 = \frac{(n+1)  \left(n(n+1)+12\right)G}{3 \rho  R_o^2} + \frac{(n+2)(n+1)(n-1)\gamma}{\rho R_o^3}.
    \end{aligned}
    \label{eq:murakami_lin_coef}
\end{equation}
\subsection{Vorticity Equation}
Taking the curl of the the momentum balance, there are non-zero terms for the $\theta$ and $\phi$ components.
Projecting each equation into their respective spherical harmonic basis, a single equation representative of the vorticity equation is
\begin{equation}
\begin{aligned}
    &3\rho\frac{R^{n+4}}{r^{n+9}}\Bigg( \epsilon_n \left(\dot{R}^2 \left((n+5) r^3-(n+7)R^3\right)+r^3R \ddot{R}\right)+r^3R\dot{R} \dot{\epsilon}_n\Bigg) =  \frac{R^{n+3}}{ \varrho^{7}r^{n+13}}\\
    & \times\Bigg( \zeta  G \Big(\alpha  \Big(\zeta  \left(4 n^2+n-177\right) r^9-\zeta ^4 (n-6) (n+11)+15 (n+3) r^{12}-8 \zeta ^2 (n (n+3)-36) r^6\\
    &+\zeta ^3 (n (5 n+21)-222) r^3\Big)-(3 \alpha -1) \varrho ^{2} r^4 \Big(3 \zeta ^2 (n+7)+5 (n+3) r^6-\zeta  (7 n+37) r^3\Big)\Big)+30  \mu  (n+3) \varrho ^{7} r^5 R^2 \dot{R}\Bigg)
    \end{aligned}
\end{equation}
where, $\zeta = R_o^3 \left(\lambda^3 - 1\right)$.

\section{Non-spherical laser-induced inertial cavitation setup}

\begin{figure}
    \centering
    \includegraphics[width=0.75\linewidth]{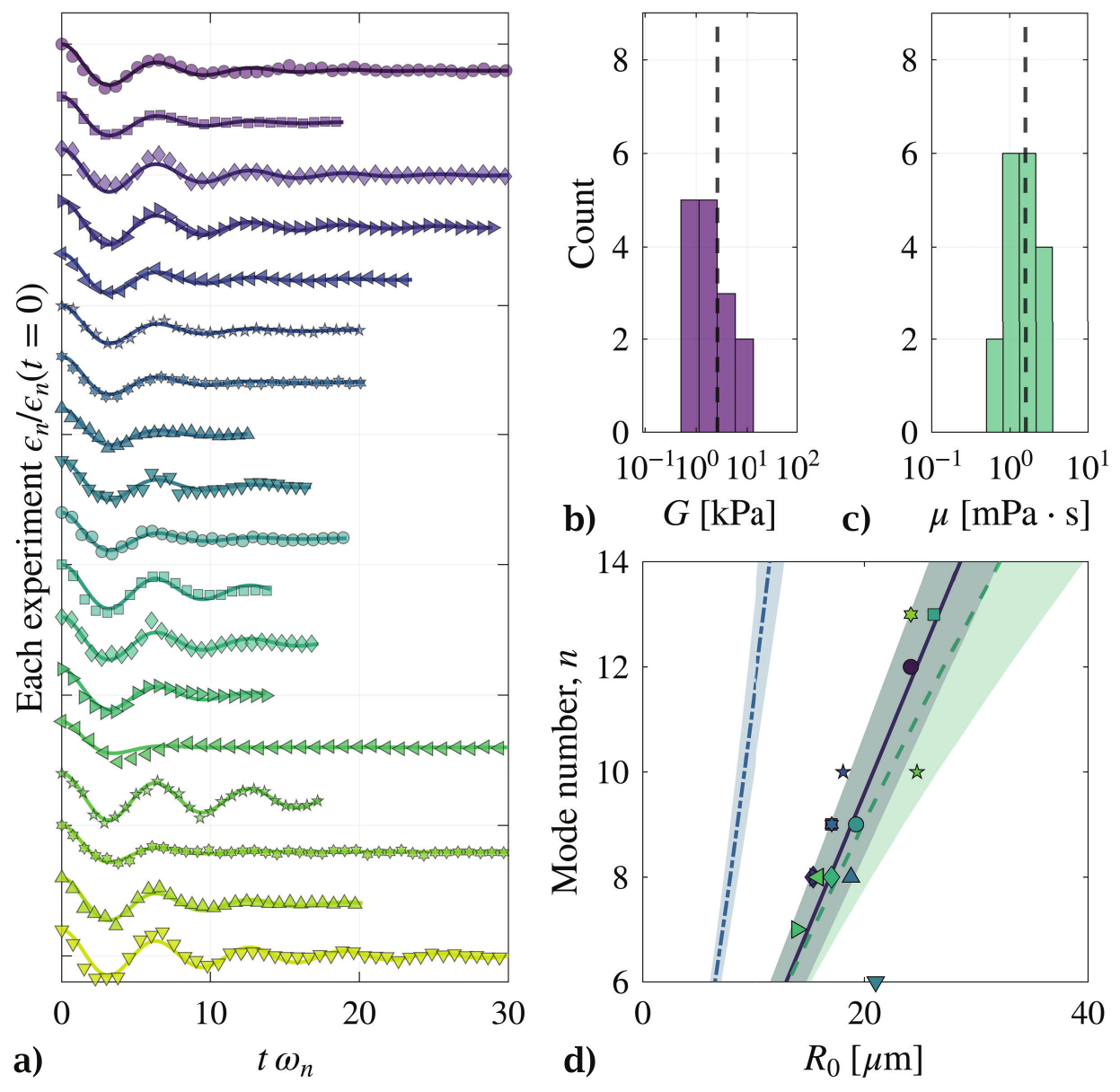}
    \caption{\textbf{a)}Evolution of the normalized amplitude for the most unstable mode for 5$\%$ gelatin, experimental data (symbols) and model prediction (lines), different colors correspond to different experiments. \textbf{b)} and \textbf{c)} Histograms of the calibrated shear modulus and viscosity respectively. 
    \textbf{d)} Most unstable mode as a function of equilibrium radius. Symbols and purple shaded regions/lines: present experimental data and current theory, respectively; green: \citet{Murakami2020} model; and steel blue: \citet{Yang2021} model. 
    Shaded regions around the mean predictions are from using the standard deviation of the distribution of extracted shear moduli.
    }
    \label{fig:gel_5_results}
\end{figure}

\begin{figure}
    \centering
    \includegraphics[width=0.35\linewidth]{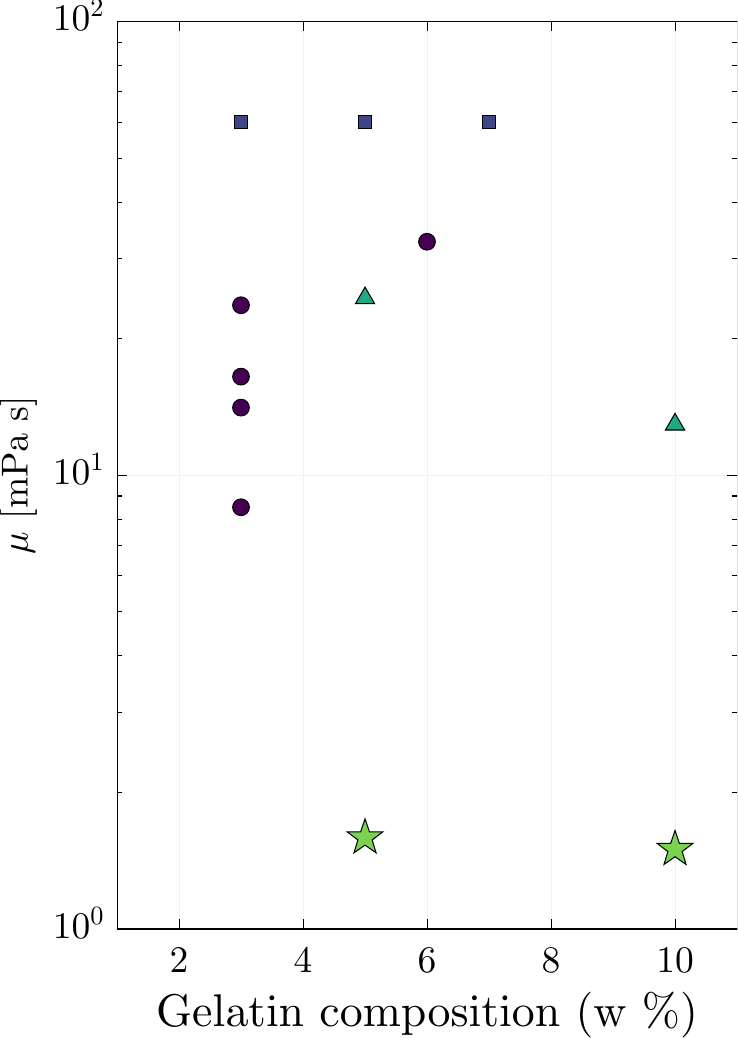}
    \caption{Shear modulus as a function of gelatin gel composition from the literature compared with present work. 
    Purple circles: \citet{Hamaguchi2015}; blue squares: \citet{Yoon2011}; green upward triangles: \citet{Abeid2024_droplet} high strain rate; and green stars: present work.} 
    \label{fig:material_props_lit}
\end{figure}

\begin{table}[htbp]
    \centering
    \caption{Summary of calibrated material properties to each experimental dataset for the ultra-sound forced experiments.}
    \label{tab:ultrasound_experiment_summary}
    \begin{tabular}{c c c c}
        \hline
        \textbf{Experiment \#} & \textbf{Gel (wt\%)} & \textbf{Elasticity}, $G$ [kPa] & \textbf{Viscosity}, $\mu$ [Pa s] \\
        \hline
        1  & 5  & 2.53 & 1.48 \\
        2  & 5  & 0.77 & 1.45\\
        3  & 5  & 0.77 & 0.91 \\
        4  & 5  & 0.99 & 0.88 \\
        5  & 5  & 3.14 & 1.52 \\
        6  & 5  & 0.50 & 1.42 \\
        7  & 5  & 0.97 & 1.64 \\
        8  & 5  & 2.86 & 2.38 \\
        9  & 5  & 12.86 & 3.48 \\
        10 & 5  & 2.22 & 1.92 \\
        11 & 5  & 3.74 & 1.05 \\
        12 & 5  & 1.86 & 1.07 \\
        13 & 5  & 1.26 & 1.24 \\
        14 & 5  & 0.50 & 2.68 \\
        15 & 5  & 1.19 & 0.50 \\
        16 & 5  & 9.64 & 3.41 \\
        17 & 5  & 0.62 & 1.01 \\
        18 & 5  & 0.50 & 0.58 \\
        \hline
        1 & 10 & 7.82 & 2.44 \\
        2 & 10 & 8.36 & 1.96\\
        3 & 10 & 18.41 & 5.04 \\
        4 & 10 & 8.16 & 0.80 \\
        5 & 10 & 4.48 & 0.33 \\
        6 & 10 & 5.72 & 0.48 \\
        7 & 10 & 5.65 & 0.51 \\
        8 & 10 & 3.89 & 0.44\\
        \hline
    \end{tabular}
\end{table}

Nonspherical bubbles were captured in the laser-induced inertial cavitation (LIC) experiments as described in \citet{YANG2020}. 
Polyacrylamide hydrogel was used with an equilibrium shear modulus of \SI{2.77}{\kilo\pascal}.
40\% acrylamide solution and 2\% bis solution (Bio-Rad, Hercules, CA) were mixed to a final concentration of 8.0\%/0.08\% Acrylamide/Bis (v/v) and cross-linked with 0.5\% Ammonium Persulfate (APS) and 1.25\% N,N,N’,N’-tetramethylethane-1,2-diamine (TEMED) (ThermoFisher Scientific, USA). 
Single microcavitation bubbles were generated in these polyacrylamide hydrogels using a \SI{6}{\nano\second} pulse from an adjustable 1-25 mJ Q-switched \SI{532}{\nano\meter} Nd:YAG laser (Continuum Minilite II, Milpitas, CA). 
Laser pulses are beam-expanded then pass through the back of a Nikon Ti:Eclipse microscope (Nikon Instruments, Long Island, NY) and fill the back aperture of a Nikon Plan Fluor 20X/0.5 NA objective.
A dichroic mirror (Semrock, Rochester, NY) was used to pass the \SI{532}{\nano\meter} beam-expanded laser light to the imaging objective and reflect the longer \SI{640}{\nano\meter} illumination light from a SI-LUX640 laser illumination system (Specialized Imaging, Pitstone, United Kingdom) into a Kirana5M high-speed camera (Specialized Imaging, Pitstone, United Kingdom).
Cavitation bubbles were recorded at \SI{2E6}{} frames per second by synchronizing a \SI{500}{\nano\second} exposure time from the camera with \SI{250}{\nano\second} illumination pulse widths from the SI-LUX illumination system before laser exposure to the sample. 
Cavitation bubbles were generated consistently at \SI{600}{\micro\meter}  above the bottom surface and separated from edges and bubbles by at least \SI{1}{\milli\meter}.

\begin{figure}
    \centering
    \includegraphics[width=0.85\linewidth]{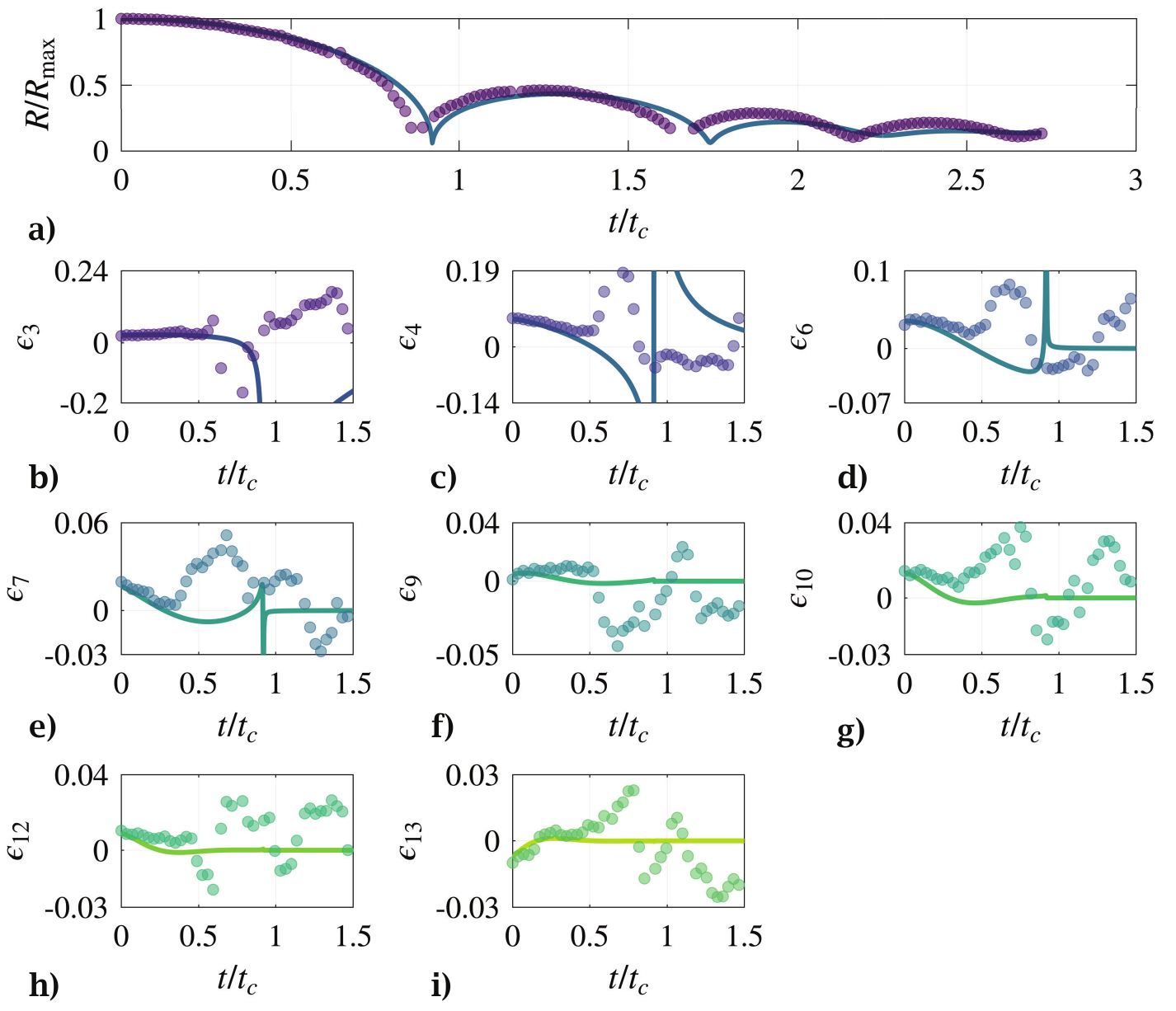}
    \caption{Experiment 1. Non-spherical laser induced bubble \textbf{a}) mean radius evolution in time, \textbf{b-i})  third, fourth, sixth, seventh, ninth, tenth, twelfth, and thirteenth mode perturbation amplitude over time. 
    Circles: experimental data; and lines: calibrated model prediction.}
    \label{fig:LIC_exp3}
\end{figure}

\begin{figure}
    \centering
    \includegraphics[width=0.85\linewidth]{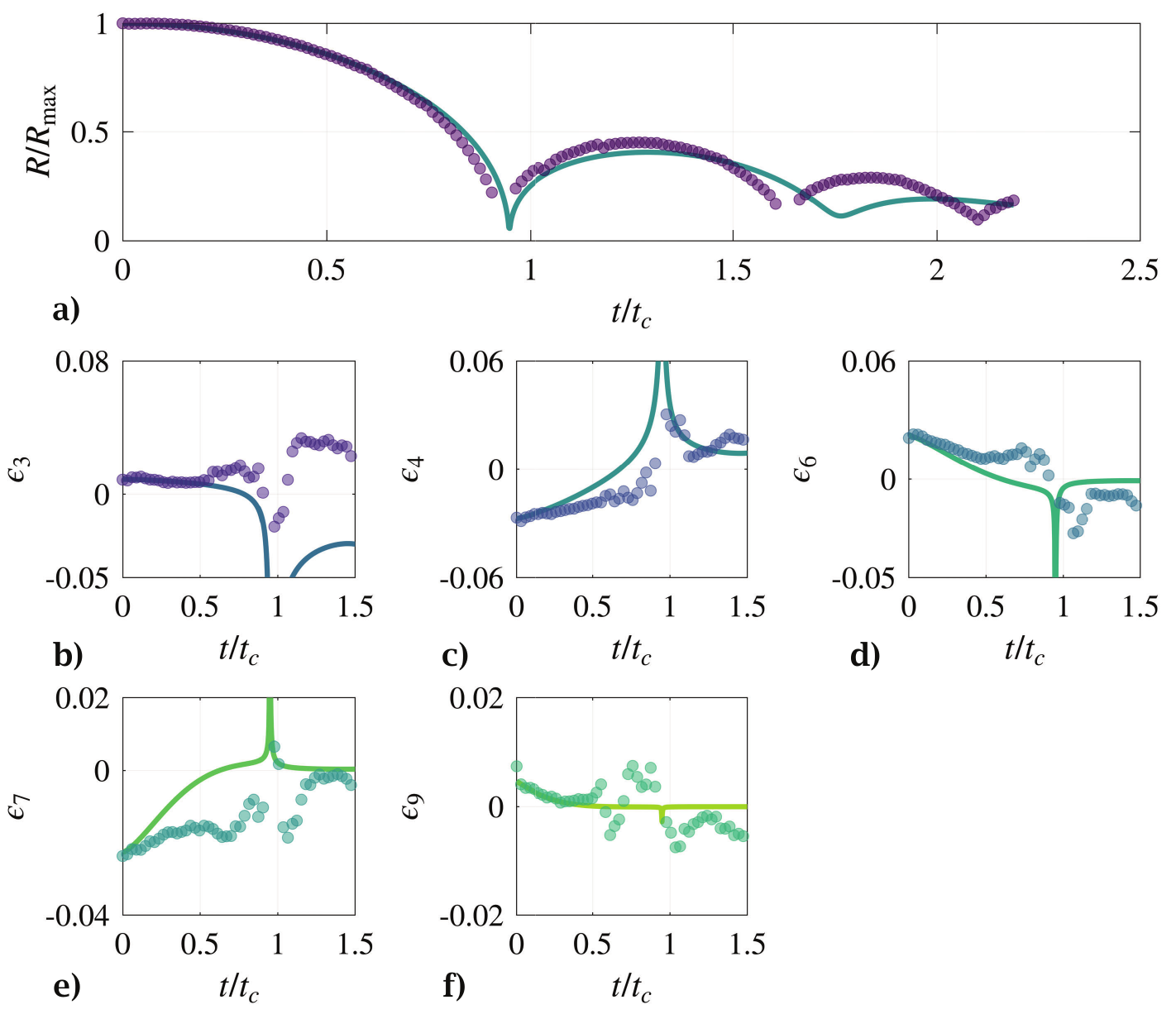}
    \caption{Experiment 2. Non-spherical laser induced bubble \textbf{a}) mean radius evolution in time, \textbf{b-f})  third, fourth, sixth, seventh, and ninth mode perturbation amplitude over time. 
    Circles: experimental data; and lines: calibrated model prediction.}
    \label{fig:LIC_exp5}
\end{figure}

\begin{figure}
    \centering
    \includegraphics[width=0.85\linewidth]{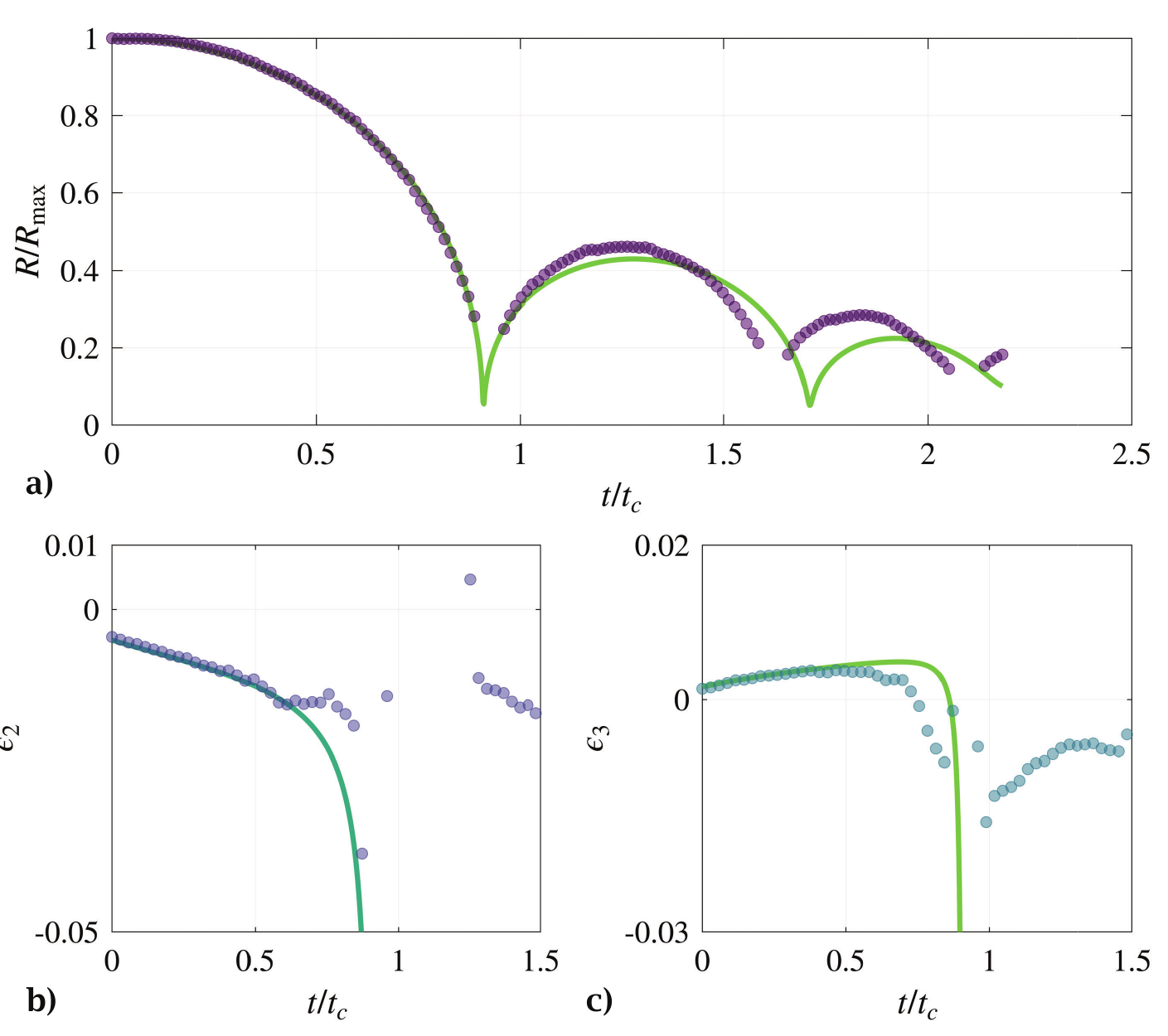}
    \caption{Experiment 3. Non-spherical laser induced bubble \textbf{a}) mean radius evolution in time, \textbf{b-c})  third, and fourth mode perturbation amplitude over time. 
    Circles: experimental data; and lines: calibrated model prediction.}
    \label{fig:LIC_exp11}
\end{figure}

\begin{table}[htbp]
    \centering
    \caption{Summary of calibrated material properties to each experimental dataset for the non-spherical LIC  experiments.}
    \label{tab:ultrasound_experiment_summary}
    \begin{tabular}{c c c c}
        \hline
        \textbf{Experiment \#} & \textbf{Strain-stiffening elasticity}, $\alpha$ [-/-] & \textbf{Viscosity}, $\mu$ [mPa$\cdot$s] \\
        \hline
        1    & 0.040 &  91.1 \\
        2    & 0.056 & 181.2 \\
        3    & 0.147 & 89.9 \\
        4 (main manuscript)  & 0.090 & 240.1 \\
        \hline
    \end{tabular}
\end{table}

\section{Data Processing}
A Fourier transform of the bubble surface data points is taken to extract the perturbation amplitudes.
For the case of ultrasound forced experiments, we assume that the bubble surface is axis-symmetric, in contrast, in the non-spherical LIC experiments we observe formation of non-axisymmetric instabilities \cite{Yang2021}.
The Fourier basis is different from the spherical harmonic basis.
We solve the linear problem corresponding to the transformation at each frame.
In the non-axisymmetric case, since we are assuming that we are imaging at a constant value of $\theta$, the bases are different by a constant factor of $(-1)^{|m|}\sqrt{2}\sqrt{\frac{2n+1}{4 \pi}\frac{(l-|m|)!}{(l+|m|)!}}P_n^m(0)$.

We consider the dominant mode to be the mode which yields the largest root mean square deviation of the perturbation amplitude in time,
\begin{equation}
    \text{RMSD}_n = \sqrt{\frac{1}{t_f-t_i}\int_{t_i}^{t_f}\left(\epsilon_n-\overline{\epsilon}_n\right)^2 \, dt}.
    \label{eq:mean_deviation}
\end{equation}
The dominant mode perturbation data is trimmed such that the initial condition coincides with a peak in the perturbation amplitude.
Signals are rejected if $| \epsilon_n(t=0)| < 0.01$, are less than one period of oscillation, and modes which are approximately zero.
Taking a Fourier transform of the signal in time, signals are rejected if they exhibit a dominant oscillation frequency which is further than 33$\%$ from half the forcing frequency (i.e., the parametric resonance condition).

A small amount of smoothing via a Savitzky-Golay filter is applied to the non-spherical LIC data to retain the signal and reduce the noise.
We include perturbation data up to the point of collapse since non-linearities in the perturbation become dominant afterwards.
A minimum norm criterion is set on the perturbation amplitude, $\|\epsilon_n\|/K^c > 10^{-3}$, where $K^c$ is the number of frames captured between the point of maximum bubble expansion and the collapse time and $\|\cdot\|$ is the Euclidean norm.

\subsection{Inverse characterization
\label{sec:Characterization}}

A global search method (\texttt{bayesopt}) hybridized with a local search method (\texttt{lsqcurvefit}) in MATLAB is used to optimize the present model to the experimental data about the material properties $\mu, \, G, $ and $\alpha$.
For large radial oscillations, $G$ is the quasi-static value, for the polyacrylamide gel in the cavitation experiments,  $G_{\infty} = \SI{2.77}{\kilo\pascal}$, and therefore is omitted from the optimization.
\texttt{bayesopt} is used to search the entire parameter space quickly.
The best performing points are used in \texttt{lsqcurvefit} for fine grained refinement. 
The set of parameters with the lowest value of the loss function are the calibrated parameters for the given experiment.

The ultrasound forced experiments contain small radial oscillations.
Thus, we use the analytical solution of the perturbation decay for the fitting process.
$\eta_{\rm l}$ depends only on viscosity, $\mu$, and $\xi_{\rm l}$ depends on $G$ and $\gamma$.
Optimizing for surface tension creates a non-unique solution for a given experiment with associated $n$ and $R_o$: an increase in $G$ and decrease in $\gamma$ (or vice versa) will lead to the same value for $\xi_{\rm l}$.
Thus, the surface tension is quantity that is set rather than optimized.
For the gelatin gels analyzed via ultrasound forcing, surface tension will be set to \SI{32}{\milli\newton\per\meter} \cite{Kikuchi_1991}.
For the polyacrylamide gels analyzed via LIC, we use the value from \cite{Yang2021} of \SI{56}{\milli\newton\per\meter}.
We sweep over three orders of magnitude for stiffness and viscosity with bounds of $[0.1,  100]$ \SI{}{\kilo\pascal} and $[0.1,  100]$ \SI{}{\milli\pascal\second}, respectively.
The loss function for the ultrasound-forced data is set to the RMSE between the experimentally obtained dominant mode perturbation amplitude as a function of time and the values predicted by the model.
After extracting the optimal parameters to fit the dataset, we computed the $R^2$ value between the predicted data and the experimental data.
We reject datasets with $R^2 < 0.75$.

The full coupled system is solved to calibrate the non-spherical LIC experimental data.
The Keller-Miksis equation is solved for the radial dynamics with the associated coupled equations for thermal and mass transport as described in \cite{Estrada2018}.
The compressive, thermal, and mass transport effects are small compared to the inertial and constitutive effects and therefore are generate $\mathscr{O}(\epsilon_n^2)$ in the perturbation dynamics.
The parameters $\alpha$ and $\mu$ are optimized through minimizing the loss function \cref{eq:loss_funct} with bounds $[10^{-3}, 10]$ and $[10^{-3}, 1]$ \SI{}{\pascal\second}, respectively.

The optimization problem involves each mode perturbation amplitude as an objective for each instance in time.
We normalize the radial data and each mode's perturbation amplitude data by its respective Euclidean norm.
To avoid multi-objective optimization, we construct a scalar loss function where the radial error is multiplied by the number of modes considered and summed with the perturbation error. 
The scalar error function is
\begin{equation}
\begin{aligned}
    \mathcal{E} = \Bigg(\sum_k^K\left[\frac{N}{\|R^{\rm exp} \|}\left(R^{\exp}(t_k)-R^{\rm sim}(t_k) \right)\right]^2 +\sum_n^N\sum_k^{K^c}  \left[ \frac{1}{\|\epsilon_n^{\rm exp} \|}\left(\epsilon_{n}^{\rm exp}(t_k) - \epsilon_{n}^{\rm sim}(t_k) \right)\right]^2\Bigg)^{1/2},
    \label{eq:loss_funct}
    \end{aligned}
\end{equation}
where superscripts $\exp$ and sim denote experimental and simulated values, respectfully.
$N$ is the total number of modes considered from the experiment, $k$ a summation over time-frames in the experiment, and $K$ the total number of frames in the experiment.
The loss function is implemented into \textit{bayesopt} and \textit{lsqcurvefit} in MATLAB.

The results from applying the optimization procedure to the LIC experimental data are shown in \cref{fig:LIC_exp3}, \cref{fig:LIC_exp5}, and \cref{fig:LIC_exp11}.
Modes (3, 4, 7, 9, 10, 12, 13), (3, 4, 6, 7), (2) from the first, second, and third experiments respectively passed the pre-filtering criteria to be included in the optimization.
We observe agreement between the dynamics of additional modes not included in the optimization.
In \cref{fig:LIC_exp5} and \cref{fig:LIC_exp11}, we also plot modes (9) and (3) for experiments two and three, respectively.
From the first experiment, the modes presented were included in the optimization.
As seen in \cref{tab:loss_funct_values}, when these added modes (not included in the optimization procedure) are used to compute the loss function, we achieve a reduction in the scalar value.
This suggests successful model calibration.

\begin{table}[]
    \centering
    \begin{tabular}{c c c }
    \hline
       \textbf{Experiment}  & \textbf{$\mathcal{E}$ from optimization}
         & \textbf{$\mathcal{E}$ with additional modes}\\
         \hline
         1 &0.275 & -  \\
         2 & 0.160 & 0.151 \\
         3 & 0.5041& 0.402 \\
         4 (main manuscript) & 0.167 & 0.156\\
         \hline
    \end{tabular}
    \caption{Summary of loss function values from model calibration to the LIC experimental data. Values are reported from the optimization procedure as well as when additional modes are included.}
    \label{tab:loss_funct_values}
\end{table}

\begin{figure}
    \centering
    \includegraphics[width=0.75\linewidth]{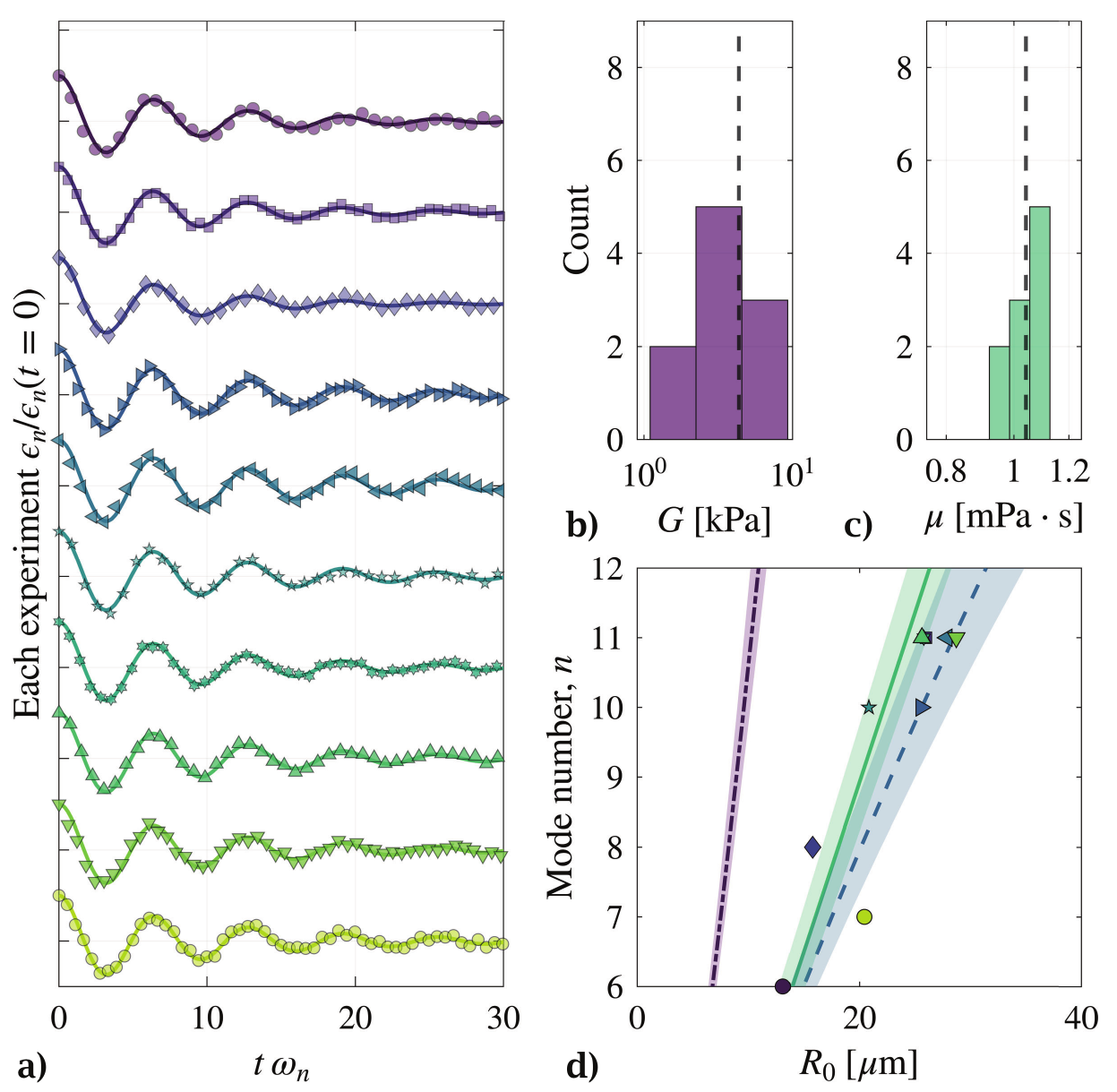}
    \caption{\textbf{a)} Evolution of the normalized amplitude for the most unstable mode for synthetic data, synthetic data (symbols) and model prediction (lines), different colors correspond to different experiments. \textbf{b)} and \textbf{c)} Histograms of the calibrated shear modulus and viscosity, respectively. 
    \textbf{d)} Most unstable mode as a function of equilibrium radius. Green symbols and shaded regions/lines: present synthetic data and current theory, respectively; steel blue: \citet{Murakami2020} model; and purple: \citet{Yang2021} model. 
    Shaded regions around the mean predictions are from using the standard deviation of the distribution of extracted shear moduli.}
    \label{fig:synth_data_raw}
\end{figure}

\begin{figure}
    \centering
    \includegraphics[width=0.75\linewidth]{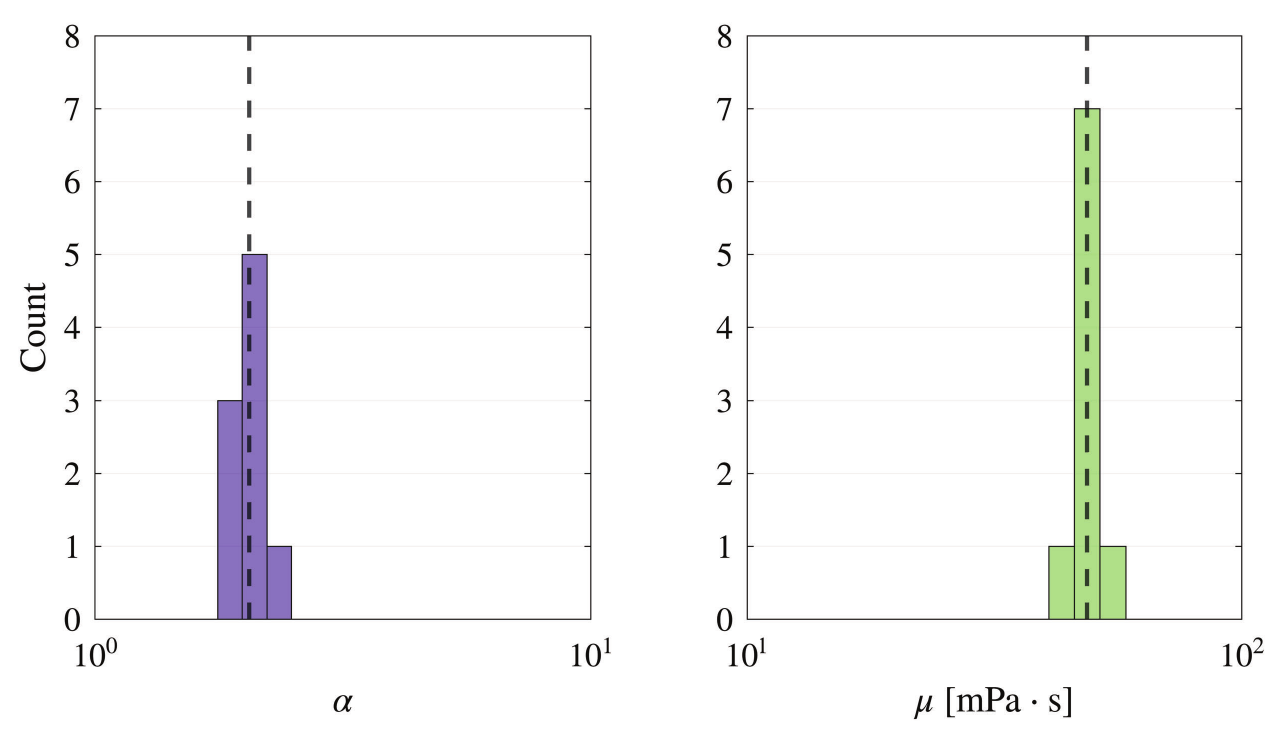}
    \caption{Material property histograms of calibrated parameters from synthetic datasets.}
    \label{fig:LIC_synt_params}
\end{figure}

\section{Inverse characterization consistency check}
We test our inverse characterization method on synthetic data that resembles the ultrasound forced datasets.
The data was generated with a shear modulus of $G = \SI{4}{\kilo\pascal}$, and a viscosity of $\mu =$ \SI{1}{\milli\pascal\second}.
Gaussian noise was injected into the perturbation amplitude at each timestep, the frequency of oscillation, and the equilibrium radius which was calculated via the parametric resonance condition for a given mode number.
The results are shown in \cref{fig:synth_data_raw}.
The mean of the extracted parameters from the datasets is $\overline{G} =$ \SI{4.37}{\kilo\pascal} and $\overline{\mu} = 1.04$ \SI{}{\milli\pascal\second}. 

Similarly, we generate noisy data resembling the LIC experiments by solving the full coupled system.
We generate 10 datasets varying the maximum radius and maximum stretch ratio as is done experimentally (data not shown).
We set the quasi-static shear modulus to $G = \SI{2}{\kilo\pascal}$, $\alpha = 2$ and $\mu = 50$ \SI{}{\milli\pascal\second}.
\Cref{fig:LIC_synt_params} shows the histogram counts of calibrated material properties.
The mean of the calibrated parameters is accurately determined and $\overline{\alpha} = 2.0$ and $\overline{\mu} = 50.11$ \SI{}{\milli\pascal\second}.
% With the level of injected noise into the synthetic data and retained accuracy on the inverse characterization, we show that calibrated values from the experimental data are accurately evaluated.

\bibliographystyle{cas-model2-names}

% Loading bibliography database
\bibliography{refs}